\documentclass[twocolumn]{aastex62}

\usepackage{verbatim}

\graphicspath{{./}{figures/}}

\received{\today}
\revised{\today}
\accepted{00-00-0000}

\submitjournal{ApJ}

\shorttitle{Multi-observing-technique study of the Be star $\omega$ CMa}
\shortauthors{Ghoreyshi et al.}

\usepackage[caption=false]{subfig}

\usepackage{soul}
\definecolor{chartreuse}{rgb}{0.87, 1.0, 0.0}
\sethlcolor{chartreuse}

 \usepackage{array,multirow,graphicx}
 \usepackage{float}

\begin{document}

\title{\Large{\bf{A Multi-observing-technique Study of the Dynamical Evolution of the Viscous Disk around the Be Star $\omega$ CMa.}}}


\defcitealias{ghoreyshi2018}{G18}

\correspondingauthor{Mohammad Reza Ghoreyshi}
\email{sghoreys@uwo.ca}

\author{Mohammad R. Ghoreyshi}
\affil{Department of Physics and Astronomy, Western University, London, ON N6A 3K7, Canada }
\affil{Instituto de Astronomia, Geof\'{i}sica e Ci\^{e}ncias Atmosf\'{e}ricas, Universidade de S\~{a}o Paulo, Rua do Matao 1226, SP 05508-900, Brazil}

\author{Alex C. Carciofi}
\affil{Instituto de Astronomia, Geof\'{i}sica e Ci\^{e}ncias Atmosf\'{e}ricas, Universidade de S\~{a}o Paulo, Rua do Matao 1226, SP 05508-900, Brazil}

\author{Carol E. Jones}
\affil{Department of Physics and Astronomy, Western University, London, ON N6A 3K7, Canada }

\author{Daniel M. Faes}
\affil{Gemini Observatory/NSF's NOIRlab, 670 N. A'ohoku Place, Hilo, Hawai'i, 96720, USA}

\author{Dietrich Baade}
\affil{European Organisation for Astronomical Research in the Southern Hemisphere (ESO), Karl-Schwarzschild-Str.\ 2, 85748 Garching bei M\"{u}nchen, Germany}

\author{Thomas Rivinius}
\affiliation{European Organisation for Astronomical Research in the Southern Hemisphere (ESO), Casilla 19001, Santiago 19, Chile}


\begin{abstract}

The observed emission lines of Be stars originate from a circumstellar Keplerian disk that are generally well explained by the Viscous Decretion Disk model. In an earlier work we performed the modeling of the full light curve of the bright Be star $\omega$ CMa \citep{ghoreyshi2018} with the 1-D time-dependent hydrodynamics code {\tt SINGLEBE} and the Monte Carlo radiative-transfer code {\tt HDUST}. We used the $V$-band light curve that probes the inner disk through four disk formation and dissipation cycles. This new study compares predictions of the same set of model parameters with time-resolved photometry from the near UV through the mid-infrared, comprehensive series of optical spectra, and optical broad-band polarimetry, that overall represent a larger volume of the disk. Qualitatively, the models reproduce the trends in the observed data due to the growth and decay of the disk. However, quantitative differences exist, e.g., an overprediction of the flux increasing with wavelength, too slow decreases in Balmer emission-line strength that are too slow during disk dissipation, and the discrepancy between the range of polarimetric data and the model. We find that a larger value of the viscosity parameter alone, or a truncated disk by a companion star, reduces these discrepancies by increasing the dissipation rate in the outer regions of the disk.

\end{abstract}



\section{Introduction}
\label{sect:intro}

Be stars are a specific subclass of main sequence B-type stars \citep{jaschek1981, collins1987} that are characterized by the presence of one or more hydrogen emission lines in their spectrum. The emission includes mainly the first members of the Balmer line series. They originate in a circumstellar environment that is in the form of an equatorial, dust-free disk that rotates in a (nearly) Keplerian fashion. The Be stars have initial masses from $\approx$\,3.5\,M$_{\odot}$ to $\approx$\,17\,M$_{\odot}$. In a statistical analysis, \cite{cranmer2005} found that they rotate moderately fast \citep[$\approx 0.47\,v_\mathrm{crit}$, usually seen in the early types; e.g., $\kappa$ CMa;][]{meilland2007b} to close to the critical rotational speed \citep[$\approx 0.95\,v_\mathrm{crit}$, usually seen in the late types; e.g., $\alpha$ Eri;][]{domiciano2012}. Due to fast rotation, the stellar equatorial material is loosely bound but an additional mechanism is required to launch the material with sufficient angular momentum (AM) to remain in orbit.

Non-radial pulsations \citep[NRP; e.g.,][]{rivinius2003, kee2016b} are a mechanism that may facilitate the release of material and, in turn, may play a role in the variability of Be stars \citep[][see Section\ref{subsect:omegacma}]{rivinius2013a, baade2017, baade2018a, semaan2018}. Changes in both brightness and spectral line profiles are typical variations in Be stars. They are known to be variable on a range of timescales from hours to years \citep{peters1986, hanuschik1993, baade2016}. Associated with the photosphere, NRPs are the cause of both short- \citep{baade2000, huat2009} and intermediate-period variability, the latter through the non-linear coupling of several NRP modes \citep{baade2018a}. Disk processes, on the other hand, cause variability on all timescales. For instance, one-armed density oscillations \citep{okazaki1997, stefl2009} usually result in variations from months to several years. The most frequent cause of disk variability is changes in the rate of AM injection from the central star into the disk, $\dot J$, which manifests itself on all timescales from days to weeks \citep[e.g.,][]{carciofi2007,levenhagen2011} to years and decades \citep{haubois2012, rimulo2018}. Finally, binarity effects are also an important source of intermediate-period disk variations \citep[ and references therein]{panoglou2018}.

Outbursts and quiescence are routinely observed states in Be stars \citep{rivinius1998b} and they are attributed to long term, secular variations in the disk. When the variations are both of lower amplitude and duration, outbursts are commonly referred to as flickers \citep[e.g.,][]{keller2002, rimulo2017, rimulo2018}. For a (nearly) pole-on system, an outburst is typically exhibited by a rapid rise in the visible and infrared emission. Outbursts are commonly associated to disk formation due to mass being ejected by the star, and the excess is caused by a larger light emitting and scattering area \citep[see][]{haubois2012}. Conversely, if the system is seen edge-on the outburst will appear as a quick decline in brightness because the cooler disk obscures part of the hotter surface of the star. Usually, an outburst is followed by a more gradual decay (or rise, in the edge-on case) back to quiescence. A quiescence phase is associated with either the cessation or reduction of AM loss and the ensuing dissipation of the disk \citep{haubois2012, ghoreyshi2018}.

The Viscous Decretion Disk (VDD) model has been successful in reproducing the observed variations of these disks \citep[e.g.,][]{carciofi2009, carciofi2010, carciofi2012, klement2015, klement2017, klement2019, faes2016, baade2018a, rimulo2018, ghoreyshi2018, dealmeida2020, suffak2020}. In the VDD model, the material ejected by the star carries AM which is redistributed within the disk facilitated by viscosity. Some material remains in orbit and slowly diffuses outward to form the disk, while most of it falls back onto the star \citep{okazaki2002}. 

For approximately steady state disks (i.e., a disk fed at a constant rate for an extended period of time) the VDD model has a straightforward solution if one assumes the disk is isothermal \citep[e.g.,][]{bjorkman1997, okazaki2001, bjorkman2005}. The first attempt to understand the dynamical evolution of circumstellar disks around isolated Be stars was done by \cite{jones2008}. This was later followed by a systematic study by \cite{haubois2012} who coupled the 1-D time-dependent hydrodynamics code {\tt SINGLEBE} \citep[][see Section~\ref{subsect:singlebe}]{okazaki2007} and the {\tt HDUST} radiative transfer code \citep[][see Section~\ref{subsect:hdust}]{carciofi2006a, carciofi2008b}.

\cite{shakura1973} introduced the $\alpha$-viscosity prescription that links the scale of the turbulence to the (vertical) scale of the disk by a constant called viscosity parameter, $\alpha$, with the following formula
\begin{equation}
\label{alpha_definition}
\nu = \frac{2}{3}\alpha c_{s}H,
\end{equation}
where $\nu$ represents the viscosity, $c_{s}$ is the isothermal sound speed and $H$ is the disk scale-height. The $\alpha$ parameter is usually assumed to be constant and it controls the timescale of disk evolution. A large $\alpha$ speeds up the diffusion process and vice versa. 

\subsection{$\omega$ Canis Majoris}
\label{subsect:omegacma}

$\omega$ (28) CMa (HD\,56139, HR\,2749; HIP\,35037; B2 IV-Ve) is one of the brightest Be stars in the sky (with $m_{\rm v} \approx$ 3.6 to 4.2~mag) and has caught attention of observers for more than five decades. It is a nearly pole-on star ($i\approx 15^{\circ}$) so the measured projected rotational velocity of 80 ${\rm km\,s^{-1}}$ \citep{slettebak1975} is only a fraction of the true equatorial velocity, estimated to be 350 ${\rm km\,s^{-1}}$ \citep{maintz2003}. A 1.37-day line-profile variability has been observed in $\omega$ CMa suggesting that it is a non-radial pulsator \citep{baade1982a}. Later, this was confirmed by \cite{stefl1999} and \cite{maintz2003} by studying the line-profile variations caused by NRP for various photospheric absorption lines of different species including Balmer lines, He\,{\sc i}, Mg\,{\sc ii}, and Fe\,{\sc ii}. Recently, from space photometry with BRITE-Constellation \citep{weiss2014}, \cite{baade2017} found that the 0.73\,d$^{-1}$ frequency (corresponding to the 1.37-d period) is part of a NRP frequency group between $\sim$0.55\,d$^{-1}$ and $\sim$0.8\, d$^{-1}$. Another frequency group between $\sim$1.15\,d$^{-1}$ and $\sim$1.45\,d$^{-1}$ seemed to exhibit a much increased amplitude at a time when the mean brightness also increased, i.e., matter was ejected into the disk when the NRP amplitude was high. Observations with the TESS satellite of hundreds of Be stars have established such a correlation in dozens of other Be stars \citep{labadie2020}, suggesting that nonlinear coupling of NRP modes \citep{baade2018b} can indeed lead to mass ejection events in Be stars. The stellar parameters of $\omega$ CMa used in this study are summarized in Table~\ref{table:stellar_parameter} and were obtained by \cite{maintz2003}.

%


\begin{table*}
    \begin{center}
        \footnotesize
        \caption{Stellar, disk, and geometrical parameters of $\omega$ CMa. $m^{*}_{\rm v}$ is the magnitude of the star during a diskless phase.}
        \begin{tabular}{@{}cccc}
            \hline
            \hline
            & Parameter & Value & reference \\
            \hline
\parbox[t]{2mm}{\multirow{16}{*}{\rotatebox[origin=c]{90}{input parameters for modeling}}}      
\parbox[t]{2mm}{\multirow{9}{*}{\rotatebox[origin=c]{90}{}}}      
\parbox[t]{2mm}{\multirow{9}{*}{\rotatebox[origin=c]{90}{star}}} 
& \\     
            & $M$ & 9.0 M$_{\odot}$ & \citealt{maintz2003} \\ 
            & $L$ & 5224 L$_{\odot}$ & \citealt{maintz2003} \\           
            & $T_{\rm pole}$ & 22000 K & \citealt{maintz2003} \\
            & $R_{\rm pole}$ & 6.0 R$_{\odot}$ & \citealt{maintz2003} \\
            & $R_{\rm eq}$ & 7.5 R$_{\odot}$ & \citealt{maintz2003} \\
            & log g$_{\rm pole}$ & 3.84 & \citealt{maintz2003} \\
            & $v_{\rm rot}$ & 350 km s$^{-1}$ & \citealt{maintz2003} \\
            & $W$ & 0.73 & \citealt{maintz2003}\\              
            \cline{2-4}
\parbox[t]{2mm}{\multirow{7}{*}{\rotatebox[origin=c]{90}{}}}            
\parbox[t]{2mm}{\multirow{7}{*}{\rotatebox[origin=c]{90}{disk}}}   
            & $\dot{M}_\mathrm{inj}$ (min) & 2.0$\times10^{-10}\mathrm{M_\odot yr^{-1}}$ & \citetalias{ghoreyshi2018} \\            
            & $\dot{M}_\mathrm{inj}$ (max) & 3.7$\times10^{-7}\mathrm{M_\odot yr^{-1}}$ & \citetalias{ghoreyshi2018} \\
            & $-\dot{J}_{*,\mathrm{std}}$ (min) & 2.9$\times10^{33}\mathrm{g\,cm^2\,s^{-2}}$ & \citetalias{ghoreyshi2018} \\            
            & $-\dot{J}_{*,\mathrm{std}}$ (max) & 5.4$\times10^{36}\mathrm{g\,cm^2\,s^{-2}}$ & \citetalias{ghoreyshi2018} \\             
            & $T_{0}$ & 13200 K & \citealt{carciofi2012} \\
            & $R_{\rm out}$ & 1000 $R_{\rm eq}$ & \citetalias{ghoreyshi2018} \\
            & \\
            \hline
\parbox[t]{2mm}{\multirow{6}{*}{\rotatebox[origin=c]{90}{other}}}            
\parbox[t]{2mm}{\multirow{6}{*}{\rotatebox[origin=c]{90}{parameters}}}  
& \\          
            & $i$ & 12--18$^{\circ}$ & this work \\
            & $v_{\rm crit}$ & 436 km s$^{-1}$& \citealt{maintz2003} \\
            & $m^{*}_{\rm v}$ & 4.22 $\pm$ 0.05 & \citetalias{ghoreyshi2018} \\
            & $d$ & 280$^{+13}_{-11}$~pc &\citealt{gaia2016, gaia2020}\\  
            & \\
            \hline
        \end{tabular}
        \label{table:stellar_parameter}
    \end{center}
\end{table*}


\cite{carciofi2012} used {\tt SINGLEBE} and {\tt HDUST} to study the disk dissipation of $\omega$ CMa between 2003 and 2008. With time-dependent models of the dissipating disk they determined the $\alpha$ parameter. Moreover, they showed that the stellar wind is not a probable mechanism for AM injection into the disk. Later, their work was followed up by \citet[][hereafter G18]{ghoreyshi2018} who presented a model of the full $V$-band light curve, spanning more than 30 years of data. The model addressed any photometric variability larger than 0.05 mag and longer than 2 months. Any shorter variability was excluded because the contribution to the total gas content of the disk is small and these are usually poorly sampled. 

\citetalias{ghoreyshi2018} showed that the VDD model could reproduce the data well.
It was determined that $\alpha$ changes during the different epochs of the disk life as $\dot J$ varies over time. They also found that the light curve could only be reproduced if quiescence phases be interpreted as a {\it reduction} of $\dot J$, rather than a complete cessation of it, as it is commonly assumed \citep[see][]{carciofi2012}.

\begin{figure*}
    \centering
    \includegraphics[width=\linewidth]{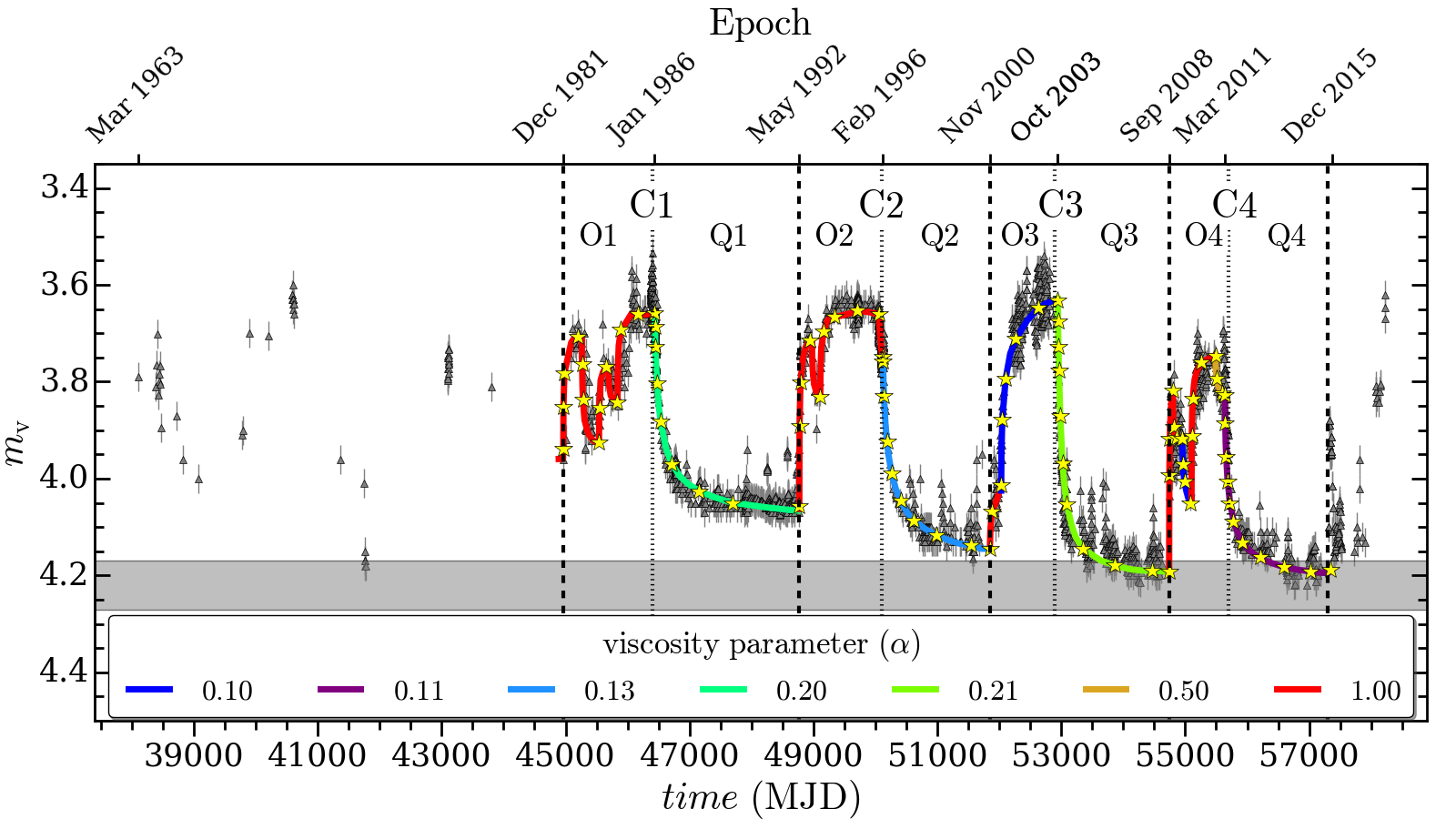}
    \caption{The model fit of the full $V$-band light curve of $\omega$ CMa obtained in \citetalias{ghoreyshi2018}. Each colored solid line illustrates an individual value for the $\alpha$ parameter, as indicated. The observed data are shown by the dark grey triangles (see the caption of Figure~1 in \citetalias{ghoreyshi2018} for references for the data). The horizontal grey band represents the visual magnitude of the star when it has no disk, determined by \citetalias{ghoreyshi2018} (see Table~\ref{table:stellar_parameter}). Selected epochs for modeling the multi-technique observations of $\omega$ CMa are marked with yellow stars (see Section~\ref{sect:multi_tech}). C$i$, O$i$ and Q$i$ stand for cycle, outburst and quiescence phases, respectively, where $i$ is the cycle number. The vertical dashed lines display the boundaries between the cycles. The vertical dotted lines show the boundaries between outburst and quiescence phases.}
    \label{fig:summary}
\end{figure*}

In \citetalias{ghoreyshi2018}, the $V$-band photometric data were investigated in detail and the results relevant to this study are summarized here. Since 1982, $\omega$ CMa exhibited quasi-regular cycles, each one lasting between about 7.0 to 10.5 years. Each cycle consists of two main parts: 1) an outburst phase represented by a fast increase in the brightness and a subsequent plateau. This increase is not always smooth, and the peak brightness plateau lasts about 2.5 to 4.0 years. 2) a quiescence phase lasting about 4.5 to 6.5 years that is characterized by a fast decline in brightness and a subsequent slow fading. During these phases the brightness of the system in the $V$ band changes from about 0$\fm$3 to 0$\fm$5. Throughout this text we refer to the cycles by C$i$ and to the phases by O$i$ and Q$i$ for outburst and quiescence\footnote{Note the difference between regular font Q for quiescence and italic font $Q$ for the Stokes parameter, and also $Q1$ and $Q3$ for the VISIR band-passes.}, respectively, where $i$ is the cycle number. All four cycles are labeled in Figure~\ref{fig:summary}.

Because the $V$-band excess comes from the innermost part of the disk \citep{carciofi2011}, the numerical solutions used in \citetalias{ghoreyshi2018} were not constrained past a few stellar radii from the star. In this paper, the same parameters as the model of \citetalias{ghoreyshi2018} are used to study other observables of $\omega$ CMa including polarimetric, spectroscopic, and photometric data at other wavelengths. Our goal is to compare the predictions of the \citetalias{ghoreyshi2018} model for a much larger volume of the disk out to 30 stellar radii. 
For this work, and following \citetalias{ghoreyshi2018}, only variability longer than two months is investigated.

Different observables emanate in different parts of the disk \citep{carciofi2011}. For example, continuum polarization originates near the star and spectral lines form in various locations based on the wavelength-dependent opacity and the source function for the line. Therefore, probing Be star disks with a variety of observational techniques and in a variety of wavelength regimes allows us to perform comprehensive studies of these systems.

The main goal of this paper is to use the VDD model to study the temporal variations seen in $\omega$ CMa's data by a variety of observational techniques. Reproducing the observed data in a wide range of wavelengths is a new challenge for the VDD model that was not previously performed.

Section~\ref{sect:omecma_obs} describes the observational data available for $\omega$ CMa. Section~\ref{sect:model} presents the theoretical concepts that were used in this work. In Section~\ref{sect:multi_tech} the modeling of the available multi-observing-technique data are presented. Finally, in Section~\ref{sect:alternative} possible solutions to solve the discrepancies between data and models are discussed. In Section~\ref{sect:conclusions} our conclusions and plans for future work are presented.


\section{Observations}
\label{sect:omecma_obs}

\begin{figure*}
    \begin{minipage}{0.5\linewidth}
        \centering
        \subfloat[]{\includegraphics[width=1.0\linewidth]{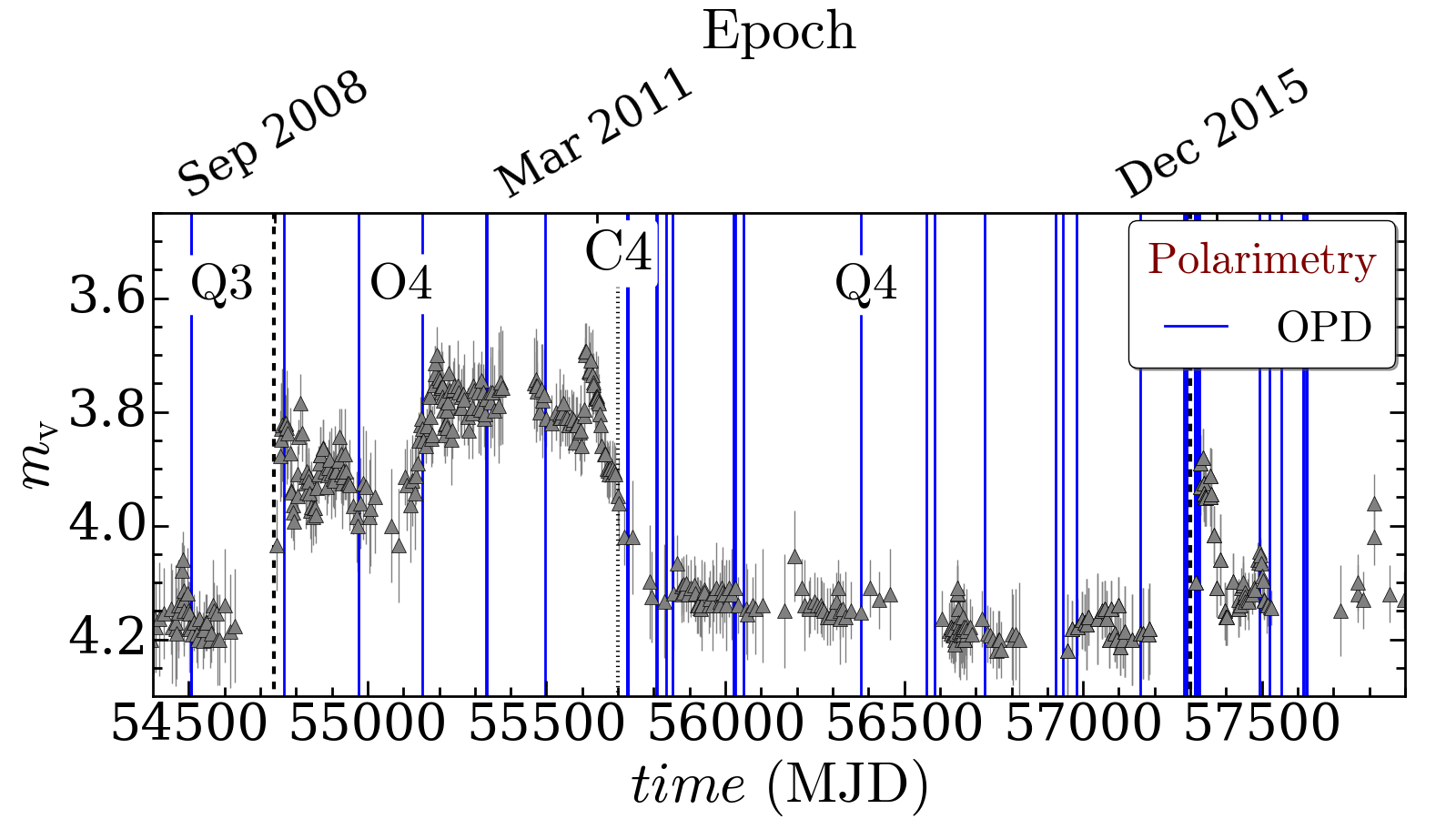}}
    \end{minipage}%
    \begin{minipage}{0.5\linewidth}
        \centering
        \subfloat[{\color{green}[S17]}]{\includegraphics[width=1.0\linewidth]{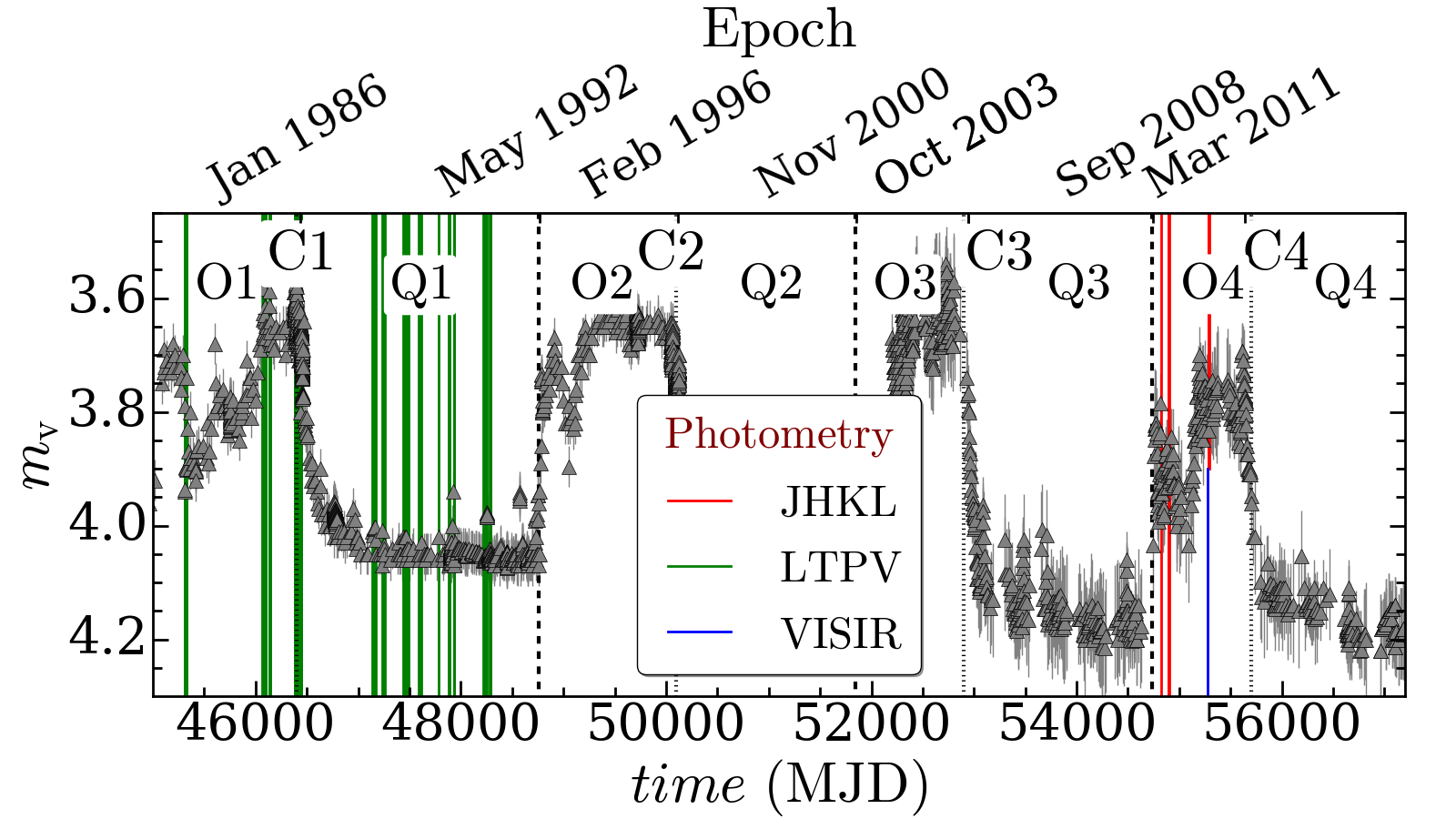}}
    \end{minipage}\par\medskip
    \begin{minipage}{1.0\linewidth}
        \centering
        \subfloat[]{\includegraphics[width=1.0\linewidth]{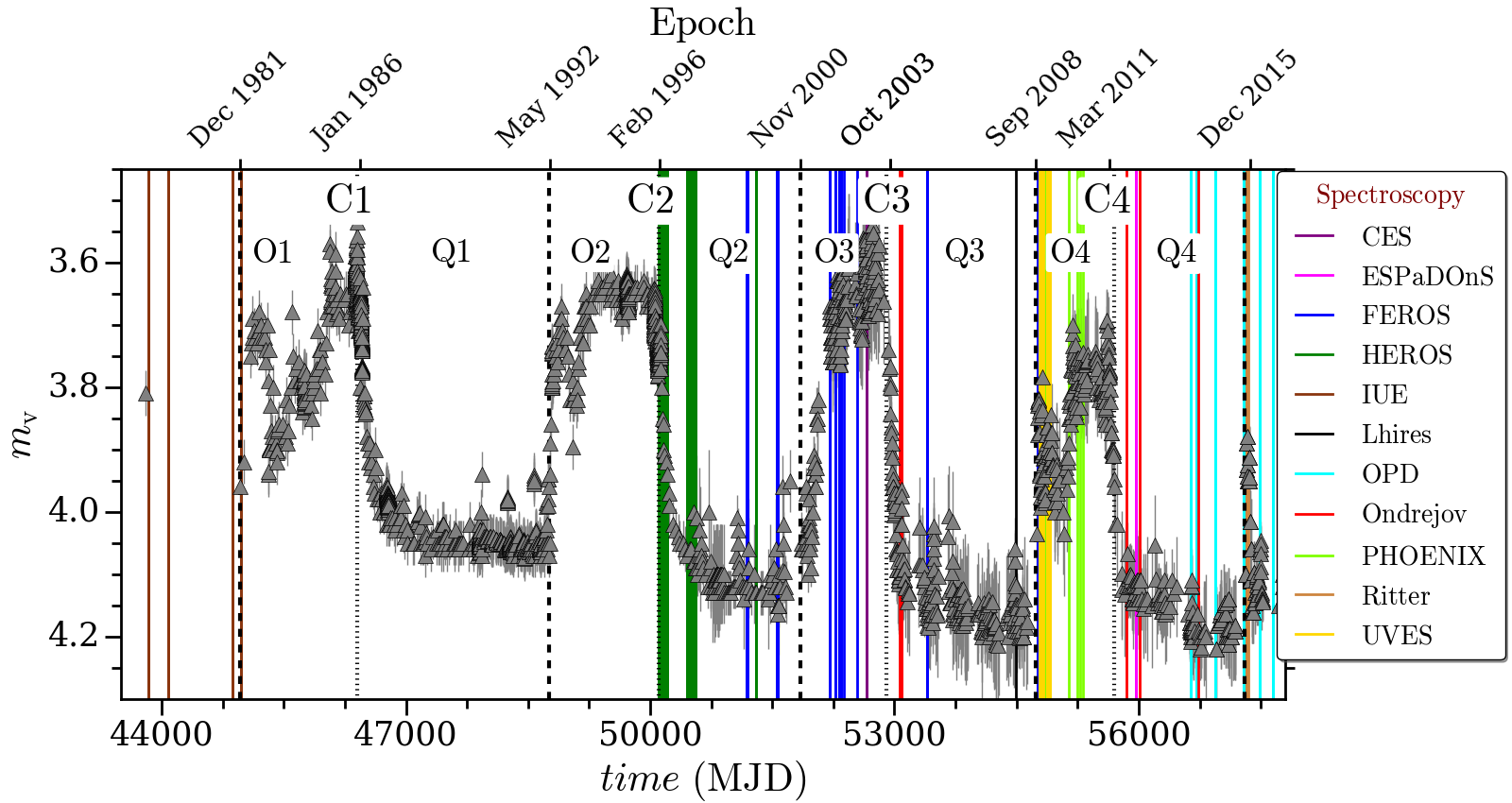}}
    \end{minipage}%
    \caption{Epochs of all available data of $\omega$ CMa in (a) polarimetry, (b) photometry, and (c) spectroscopy. Each vertical solid line represents the epoch when the data were observed. The instruments, observatories, or the photometric bands are indicated in the legend and defined in the text. The dark grey triangles in each panel display the observed $V$-band photometric data as a reference. C$i$, O$i$ and Q$i$ stand for cycle, outburst and quiescence phases, respectively, where $i$ is the cycle number. The vertical dashed lines display the boundaries between the cycles. The vertical dotted lines show the boundaries between outburst and quiescence phases.}
    \label{fig:data_dist}
\end{figure*}

A wealth of data from different observational techniques has been collected since 1963. The rich dataset covering the most recent outburst with several different techniques is an important addition for this paper. In the following we present a summary of the observed data by each technique. The epochs of all available observations are shown in Figure~\ref{fig:data_dist}. We note that the data that were observed prior to C1 (December 1981) or after C4 (December 2015)  were not included in our analysis but are shown in Figure~\ref{fig:data_dist} for completeness.


\subsection {Photometry}
\label{subsect:obs_photo}



At the end of 2008, Sebastian Otero alerted the community (in a private communication to our deceased colleague Stanislav \v{S}tefl) that a new outburst had begun, thus a broad suite of observations was undertaken. In addition to visual photometry, \textit{JHKL} photometry were obtained with the Mk II photometer  \citep{glass1973} of SAAO (South African Astronomical Observatory) and the CAIN-II (CAmara INfrarroja) Tenerife/TCS (Telescopio Carlos S\'{a}nchez) camera \citep{cabrera2006}, $Q1$- and $Q3$-band measurements were made with VISIR \citep[VLT Imager and Spectrometer for mid-InfraRed;][]{lagage2004} on the VLT/ESO (Very Large Telescope/European Southern Observatory). Finally, we also included photometric data in the {\it UBV} Johnson and {\it uvby} Str\"{o}mgren \citep{stromgren1956, crawford1958} filters  collected in the Long-Term Photometry of Variables (LTPV) project \citep{manfroid1995} during C1.

\subsection {Spectroscopy}
\label{subsect:obs_spectro}

The above campaign also produced optical echelle spectra from UVES \citep[Ultraviolet and Visual Echelle Spectrograph;][]{dekker2000} and VLT (Oct 2008-Mar 2009), {\sc feros} (Fiber-fed Extended Range Optical Spectrograph)/La Silla \citep{kaufer1999} and the 1.6m telescope at Observat\'{o}rio Pico dos Dias (OPD; Jan 2009 - 20164ri) initially using the ECASS spectrograph\footnote{This Cassegrain spectrograph consists of a 600 groove~mm$^{-1}$ grating blazed at 6563 \AA{} at the first order, resulting in a reciprocal dispersion of 1.0~\AA{}\,pixel$^{-1}$.} and, since 2012, the MUSICOS spectrograph (\citealt{bau92}).

In addition to the observational effort described above for C4 (see Figure~\ref{fig:summary}), we obtained other data for some of the previous cycles in the literature. For C1 and C2, we acquired spectroscopy from the Short-Wavelength Prime (SWP) camera of IUE\footnote{\href{https://archive.stsci.edu/iue/}{https://archive.stsci.edu/iue/}} (International Ultraviolet Explorer) and {\sc heros} (Heidelberg Extended Range Optical Spectrograph\footnote{\href{www.lsw.uni-heidelberg.de/projects/instrumentation/Heros/}{www.lsw.uni-heidelberg.de/projects/instrumentation/Heros/}})/{\sc feros}, respectively. For C3, we found spectroscopy from the CES (Coud\'{e} Echelle Spectrometer\footnote{\href{www.eso.org/public/teles-instr/lasilla/coude/ces/}{www.eso.org/public/teles-instr/lasilla/coude/ces/}}), {\sc feros}, Lhires spectroscope in Observatoire Paysages du Pilat\footnote{\href{www.parc-naturel-pilat.fr/nos-actions/architecture-urbanisme-paysage/observatoire-du-paysage/}{www.parc-naturel-pilat.fr/nos-actions/\\architecture-urbanisme-paysage/observatoire-du-paysage/}}, and Ondrejov Observatory\footnote{\href{stelweb.asu.cas.cz/web/index.php?pg=2m_telescope}{stelweb.asu.cas.cz/web/index.php?pg=2mtelescope}}. Additional spectroscopic data for C4 came from ESPaDOnS \citep[Echelle SpectroPolarimetric Device for the Observation of Stars;][]{donati2003}, OPD, PHOENIX \citep{hinkle1998}, Ritter Observatory\footnote{\href{www.utoledo.edu/nsm/rpbo/}{www.utoledo.edu/nsm/rpbo/}}, and UVES. 

\begin{figure*}
    \centering
    \includegraphics[width=0.85\linewidth]{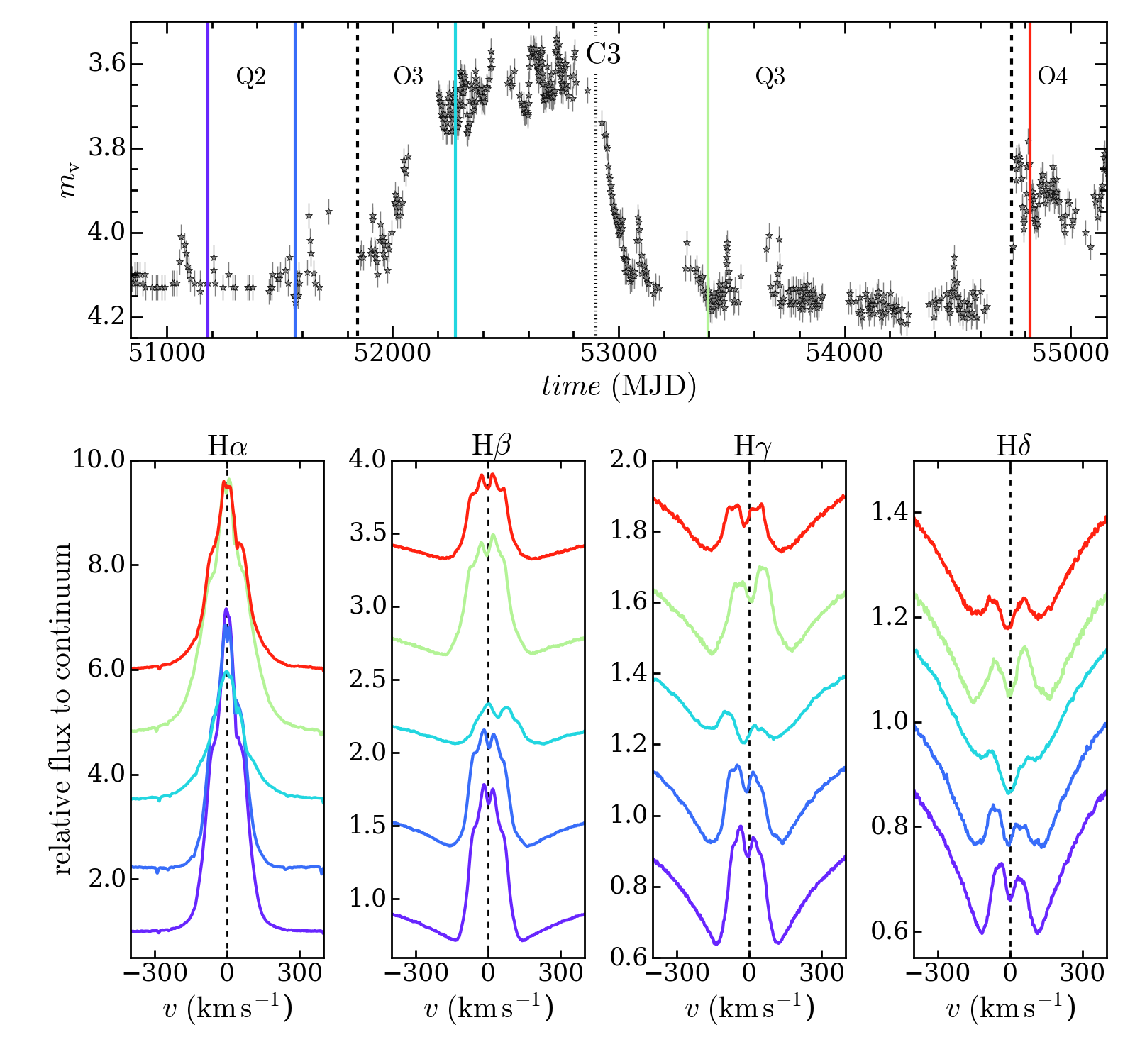}
    \caption{{\it Top panel}: The $V$-band light curve with vertical colored solid lines marking the dates the spectra shown in the bottom panel were taken. {\it Bottom panel}: Continuum normalized hydrogen lines of $\omega$ CMa observed by {\sc feros} using the same color scheme as the upper panel. The spectra were arbitrary shifted vertically for ease of comparison. C$i$, O$i$ and Q$i$ stand for cycle, outburst and quiescence phases, respectively, where $i$ is the cycle number. The vertical dashed lines display the boundaries between the cycles. The vertical dotted lines show the boundaries between outburst and quiescence phases.}
    \label{fig:feros}
\end{figure*}

Figure~\ref{fig:feros} provides an example of the observed hydrogen lines of $\omega$ CMa. The top panel shows the $V$-band photometric data with the date the spectroscopic data were observed, indicated by the colored vertical solid line. The bottom panels show the flux relative to the local continuum for the four main hydrogen lines. Usually the peak emission to the continuum ratio (E/C) of the H$\alpha$ and H$\beta$ lines is largest at the end of quiescence, and lower during the outburst. This seemingly contradictory behavior is well explained by the models as will be seen in Section~\ref{subsect:spectroscopy}. The observational logs of the spectroscopic and polarimetric data used in this paper are listed in Appendix~\ref{ap:obs_log}, and in Tables~\ref{table:spec_data_log} and \ref{table:pol_data_log}, respectively. Also, the data are available at CDS\footnote{\href{http://cdsarc.u-strasbg.fr/viz-bin/qcat?J/ApJ}{http://cdsarc.u-strasbg.fr/viz-bin/qcat?J/ApJ}}.


\subsection {Polarimetry}
\label{subsect:obs_polar}

\begin{figure}
    \centering
    \includegraphics[width=\columnwidth]{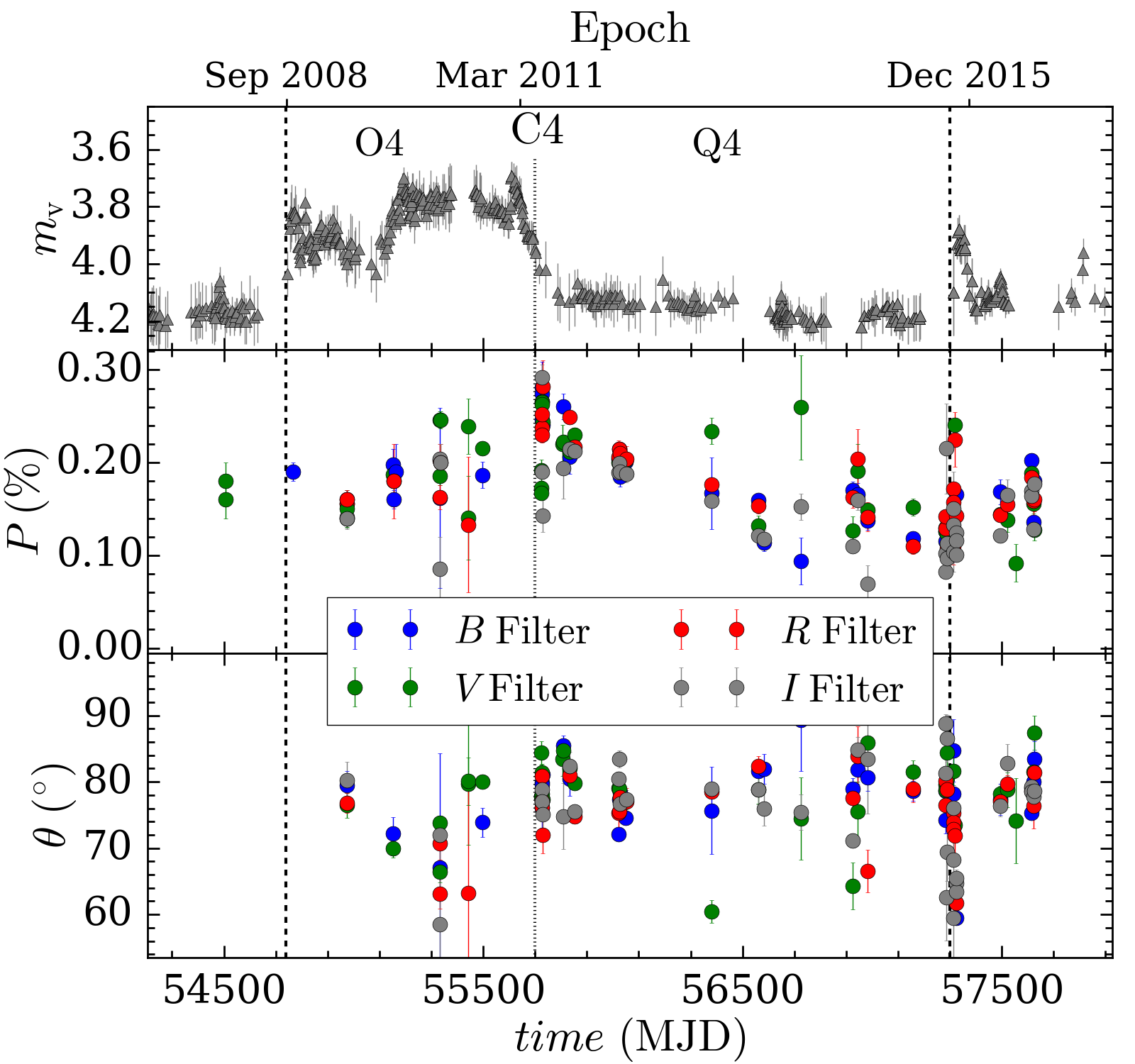}
    \caption{{\it Top panel}: The $V$-band photometric data of $\omega$ CMa. {\it Middle panel}: Polarization level in the fourth cycle in the \textit{BVRI} bands. {\it Bottom panel}: The observed position angle. C4, O4 and Q4 stand for the fourth cycle, outburst and quiescence phases, respectively. The vertical dashed lines display the boundaries between the cycles. The vertical dotted lines show the boundaries between outburst and quiescence phases.}
    \label{fig:pol_BVRI}
\end{figure}

\textit{BVRI} imaging polarimetry was obtained with the 0.6-m telescope at OPD \citep{magalhaes2006}. Reduction of the OPD polarimetric data, observed by the IAGPOL instrument, followed standard procedures outlined by \cite{magalhaes1984, magalhaes1996} and \cite{carciofi2007}. The IAGPOL  has an instrumental polarization smaller than about 0.005\% \citep{carciofi2007}. The middle panel of Figure~\ref{fig:pol_BVRI} displays the polarimetric data of C4 of $\omega$ CMa in $B$, $V$, $R$, and $I$ filters, alongside the $V$-band photometric data (top panel) and polarization angle, $\theta$, measured east from celestial north (bottom panel). The polarization level is very small, as expected by the fact that $\omega$ CMa is observed at rather small inclination angles \citep{halonen2013}. This happens because at nearly pole-on orientations an axi-symmetric disk will appear as an almost circular structure that results in the cancellation of the polarization vectors. The small observed polarization level indicates that the interstellar (IS) polarization is likely also very small. Both combined factors makes analysis of the polarization data uncertain, as will be discussed in Section~\ref{subsect:polarimetry}. 


\section{Model description}
\label{sect:model}

The principal properties of the VDD model are outlined here, and the main methods and approximations used to obtain the solutions are described in \citetalias{ghoreyshi2018}. In this model, the central star is located at the origin of a cylindrical coordinate system whose vertical axis is parallel to the rotational axis of the star ($z$ direction). The star is oblate with equatorial and polar radii  $R_{\rm eq}$ and $R_{\rm pole}$, respectively, polar temperature $T_{\rm pole}$, mass $M$, luminosity $L$, and rotational velocity $v_{\rm rot}$ that is a fraction of the critical velocity $v_{\rm crit}$. The values adopted for these parameters in this work are presented in Table~\ref{table:stellar_parameter}. The disk is assumed to lie in the equatorial plane of the star. The rotational velocities of the star and the disk are vectors in the azimuthal direction, $\phi$. Also, the disk has a radial velocity component, $v_r$, that can be negative (i.e., inflow) or positive (i.e., outflow). The most important parameters describing the disk are the AM flux injected into the disk from the star at the steady-state limit ($\dot{J}_{*,\mathrm{std}}$), the disk base temperature, $T_{0}$, and the outer radius of the disk, $R_{\rm out}$. We note that an alternative way of describing the disk feeding mechanism uses the mass injection rate into the disk, $\dot{M}_{\mathrm{inj}}$ which is related to $\dot{J}_{*,\mathrm{std}}$ by
\begin{equation}
\label{jdot_steady}
-\dot{J}_{*,\mathrm{std}}=\Lambda \left(GM_*R_\mathrm{eq}\right)^\frac{1}{2}\dot{M}_\mathrm{inj}\left(\bar{r}_\mathrm{inj}^\frac{1}{2}-1\right)\,,
\end{equation}
where $\Lambda$ is a dimensionless quantity greater than 1 and is given by $\Lambda=1/(1-\bar{r}_\mathrm{out}^{-\frac{1}{2}})$, with $\bar{r}_\mathrm{out}=R_\mathrm{out}/R_\mathrm{eq}$ and $R_\mathrm{out}\gg R_\mathrm{eq}$ \citep{rimulo2018}. One of the reasons that we do not investigate $\dot{M}_{\mathrm{inj}}$ in this paper is because it cannot be determined observationally. More details and discussion about the relationship between $\dot{M}_{\mathrm{inj}}$ and $\dot{J}_{*,\mathrm{std}}$ can be found in \cite{rimulo2018} and \citetalias{ghoreyshi2018}. The magnitude of $\dot{M}_{\mathrm{inj}}$ alongside the values of the parameters used for modeling are listed in Table~\ref{table:stellar_parameter}.

The calculations presented here were mainly completed by two computational codes: the 1D time-dependent hydrodynamics code {\tt SINGLEBE} and {\tt HDUST}. In the following the codes are briefly introduced.


\subsection {The \textsc{SINGLEBE} code}
\label{subsect:singlebe}

{\tt SINGLEBE} solves the isothermal 1D time-dependent fluid equations \citep{pringle1981} in the thin disk approximation, and provides the disk surface density, $\Sigma (r, t)$. 

The 1D grid used in the {\tt SINGLEBE} code models the disk between $R_{\rm eq}$ (the equatorial radius of the star) and $R_{\rm out}$ (the outer radius of the disk). The grid is a logarithmic array with an optional number of cells. One cell is arbitrary selected as the location where mass from the central star is entirely injected, $\bar{r}_\mathrm{inj}R_\mathrm{eq}$, where $\bar{r}_\mathrm{inj}$ is a dimensionless quantity. The viscosity parameter, as a function of time, $\alpha(t)$, and $\dot{J}_{*,\mathrm{std}}(t)$ are the input parameters of the code. {\tt SINGLEBE} determines how the injected matter spreads in the disk.
More details about {\tt SINGLEBE} can be obtained in the original publication \citep{okazaki2007}, and a description of the boundary conditions adopted can be found in \cite{rimulo2018}.


\subsection {\textsc{HDUST} code}
\label{subsect:hdust}

The Monte Carlo radiative transfer code {\tt HDUST} is a fully three-dimensional (3D) code that simultaneously solves the radiative equilibrium, the radiative transfer, and non-LTE statistical equilibrium equations to obtain the ionization fraction, hydrogen level populations, and electron temperature as a function of position in a 3D envelope around the star \citep{carciofi2004, carciofi2006a, carciofi2008b}. In order to convert the surface density provided by {\tt SINGLEBE} to volume density, {\tt HDUST} uses a Gaussian vertical density profile with a 1.5 power-law isothermal disk scale height. With these quantities, {\tt HDUST} produces the emergent spectral energy distribution (SED), including emission line profiles, as well as the polarized spectrum and synthetic images. 

To date, {\tt HDUST} has been used in a variety of theoretical studies of Be stars disks \citep[e.g.,][]{carciofi2006a, carciofi2008b, haubois2012, haubois2014, faes2013}. The most relevant study for this work is \cite{haubois2012} who used the hydrodynamic simulations to show the capability of the VDD model for reproducing the light curves of Be stars. In addition, {\tt HDUST} was used in several studies where the model predictions were constrained by observations such as visible and infrared photometry \citep[e.g.,][]{carciofi2012, baade2018a, rimulo2018, ghoreyshi2018}, radio photometry \citep[e.g.,][]{klement2017, klement2019}, polarimetry \citep[e.g.,][]{carciofi2007, carciofi2009, faes2016}, spectroscopy \citep[e.g.,][]{carciofi2010, suffak2020}, and spectro-interferometry \citep[e.g.,][]{carciofi2009,klement2015, faes2016, dealmeida2020}.

As mentioned earlier, the central star is oblate. Consequently, the polar regions of the star have a greater effective gravity than the equatorial regions. According to the \cite{vonzeipel1924} theorem, this latitudinal dependence of the effective gravity causes a latitudinal dependence of the flux, in the sense that the poles are brighter (hotter) and the equator darker (cooler). This gravity darkening effect plays a key role in determining the surface distribution of the flux, and therefore the latitudinal dependence of the temperature.

In its original formalism, applicable for a purely radiative envelope, the von Zeipel theorem can be written as 
\begin{equation}
    T_{\rm eff}(\theta) \propto g_{\rm eff}^{\beta}(\theta)\,,
    \label{eq:vonzeipel}
\end{equation}
with $\beta = 0.25$. However, interferometric studies of stellar spectral classes of F to B suggested that the $\beta$ parameter is typically in the range of 0.18--0.25 \citep{vanbelle2006, vanbelle2012, monnier2007, che2011, hadjara2018, domiciano2018} with the most likely value of 0.21 \citep{vanbelle2012}. The theoretical study of \cite{espinosa2011} suggested that the value of $\beta$ is a function of the rotational rate of the star. Following these authors, \citetalias{ghoreyshi2018} adopted a $\beta$ of 0.19 for $\omega$ CMa. The same value was used in this paper. Using the $\beta$ parameter, the polar radius of the star ($R_\mathrm{pole}$), and the critical fraction of the rotational velocity \citep[$W$, see][for how this parameter is defined]{rivinius2003} as input parameters, {\tt HDUST} calculates the geometrical oblateness and gravitational darkening of the star. 

Although {\tt HDUST} has the ability to take into account the opacity of dust grains (e.g., for B[e] stars, \citealt{carciofi2010}), our models were calculated for dust-free gaseous disks consistent with our current understanding of Be stars.


\section{Multi-observing-technique Modeling}
\label{sect:multi_tech}

Recall, in \citetalias{ghoreyshi2018} our model was limited to the $V$-band which in turn is sensitive only to variations in the disk regions very close to the star (see Figure~1 of \citealt{carciofi2011}). An important next step consists of extending the analysis to the other observables (photometry at longer wavelengths, polarimetry and spectroscopy), which probe different disk regions.

Here, various line profiles and the entire emergent polarized spectrum from the UV to the mid-infrared for about 80 selected epochs covering different phases of the disk evolution were computed using {\tt HDUST}. The selected epochs are denoted with yellow stars in Figure~\ref{fig:summary}. It is substantial to note that we do not have simultaneous observations for all the considered techniques and wavelength regions. This means there are some epochs for which only one type of observed data (for example, only spectroscopic data at the beginning of Q2) is available.



\subsection{Polarimetry}
\label{subsect:polarimetry}

The linear polarization level can be expressed in terms of the Stokes parameters, $Q$ and $U$ \citep{clarke2010} as
\begin{equation}
P = \sqrt{Q^2+U^2} \,.
\label{eq:p}
\end{equation}
The polarization position angle is
\begin{equation}
\theta = \frac{1}{2} \arctan \left(\frac{U}{Q} \right) \,.
\label{eq:theta}
\end{equation}

\begin{figure}
\begin{minipage}{\linewidth}
\centering
\subfloat[]{\includegraphics[width=0.6\linewidth]{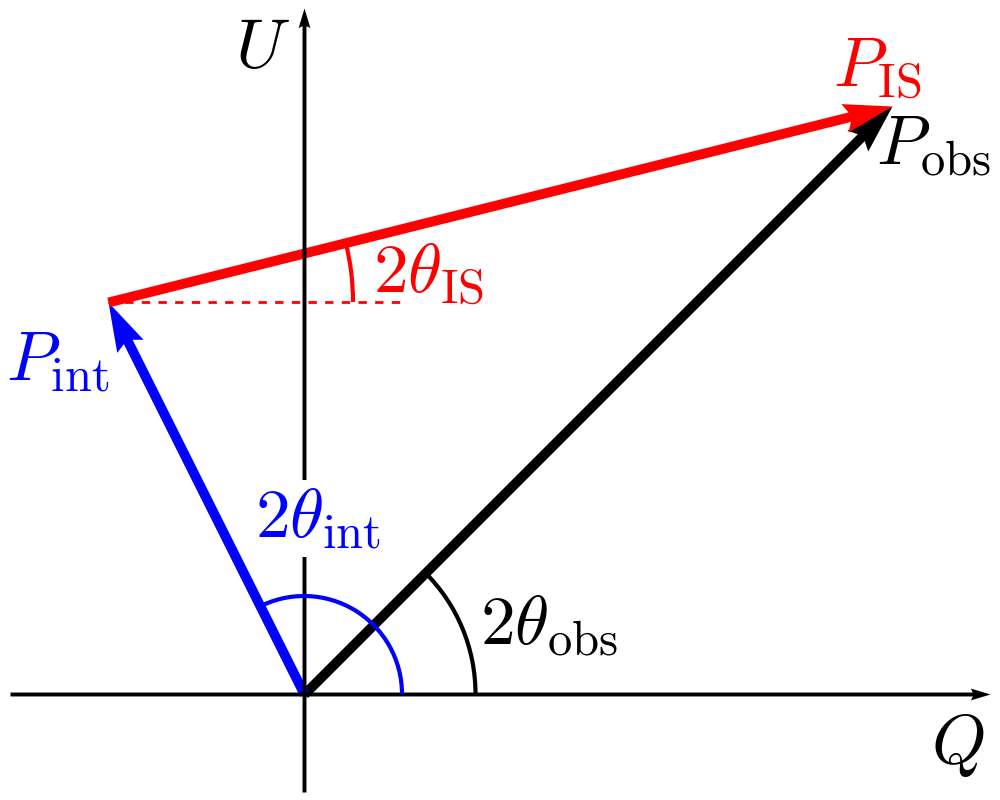}}
\end{minipage}%
\begin{minipage}{\linewidth}
\centering
\subfloat[]{\includegraphics[width=0.6\linewidth]{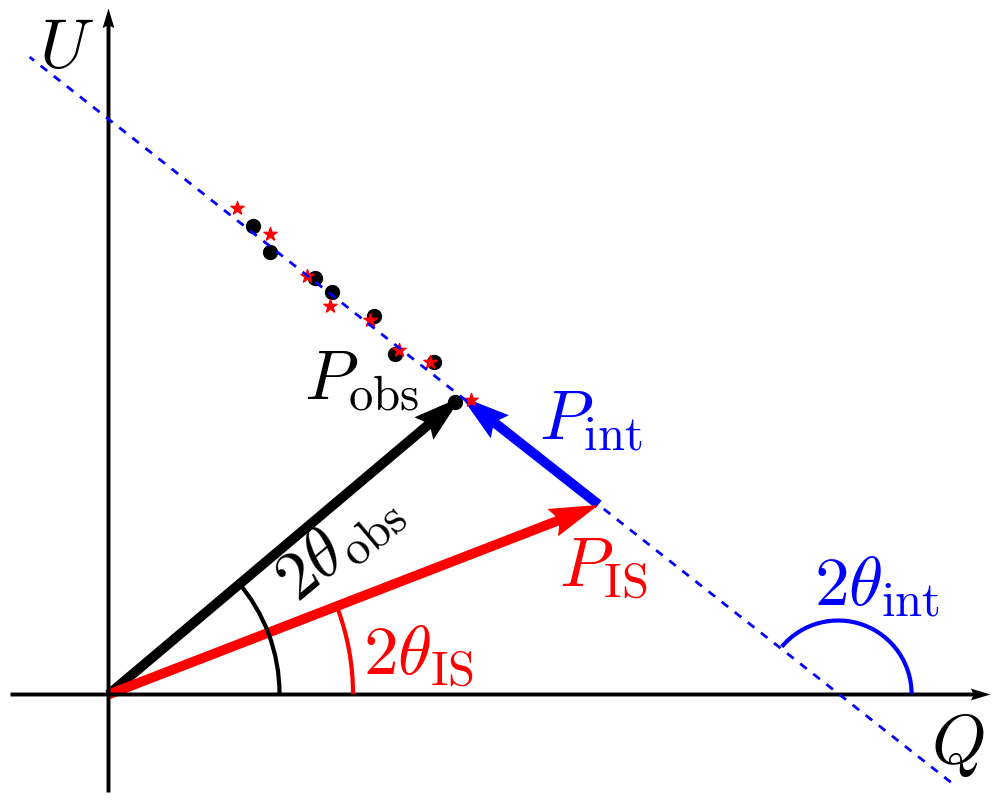}}
\end{minipage}
\caption{Schematic diagrams showing the components and the temporal variability of the observed polarization. Note that no actual data for $\omega$ CMa have been presented here. (a) The components of the observed polarization vector ($P_{\rm obs}$) in the $Q-U$ diagram: $P_{\rm IS}$ and $P_{\rm int}$. $\theta_{\rm obs}$, $\theta_{\rm IS}$ and $\theta_{\rm int}$ represent the observed, IS and intrinsic polarization angle, respectively. (b) The variations of the polarization vector and its components with the disk growth and decay. The red stars and black circles show the schematic observed polarization level during disk growth (toward up and left) and decay (toward down and right), respectively.}
\label{fig:bedna_pol}
\end{figure}

One common issue regarding interpretation of polarimetric data is the removal of the IS contribution to the observed signal. The observed polarization, decomposed in its Stokes $Q$ and $U$ parameters, can be written as:
\begin{equation}
    Q_{\rm obs} = Q_{\rm IS} + Q_{\rm int}\,,
    \label{eq:q_obs}
\end{equation}
and
\begin{equation}
    U_{\rm obs} = U_{\rm IS} + U_{\rm int}\,,
    \label{eq:u_obs}
\end{equation}
i.e., without knowing the IS components ($Q_{\rm IS}$ and $U_{\rm IS}$) of the observed polarization, the components ($Q_{\rm int}$ and $U_{\rm int}$) of intrinsic polarization 
are unknown. This is shown schematically in the top panel of Figure~\ref{fig:bedna_pol}. Measuring the IS component of the polarization can be a challenging task \citep[e.g.,][]{wisniewski2010}. As there is no reliable information about this quantity for $\omega$ CMa in the literature, we employ below three different methods, to determine the position angle of the intrinsic polarization and to estimate the IS polarization ($\theta_{\rm IS}$ and $P_{\rm IS}$).


\subsubsection{$Q-U$ method}
\label{subsubsect:qu}

The first method explores the fact that the intrinsic polarization is variable, while on the same timescale and at a given wavelength, the $P_{\rm IS}$ is not. We begin examining the bottom panel of Figure~\ref{fig:bedna_pol}, that shows, in a schematic way, how the process of formation and dissipation of a Be disk appears in the $Q-U$ diagram. The intrinsic polarization angle on the sky is $\theta_{\rm int}$. When $P_{\rm int}$ is zero (no disk), the observed polarization will be due solely to the IS component ($P_{\rm obs} = P_{\rm IS}$). As the disk grows (and dissipates), the magnitude of $P_{\rm int}$ changes, but not the angle (assuming that the disk is axi-symmetric and lies along equatorial plane). This is shown in the bottom panel of Figure~\ref{fig:bedna_pol} as the track of points along the $2 \theta_{\rm int}$ direction, which indicates that the angle of the track is a measure of $\theta_{\rm int}$ \citep{draper2014}. More specifically, $\theta_{\rm int}$ should be parallel to the minor elongation axis of the Be disk (we note that the disk may not be elliptic but appears as an ellipse in the plane of the sky in the line of sight of observer, if it is not seen pole-on). In the case of disks confined to the equatorial plane, $\theta_{\rm int}$ also describes the position angle of the spin axis of the star, measured east from celestial north.


\begin{figure}
    \centering
    \includegraphics[width=\columnwidth]{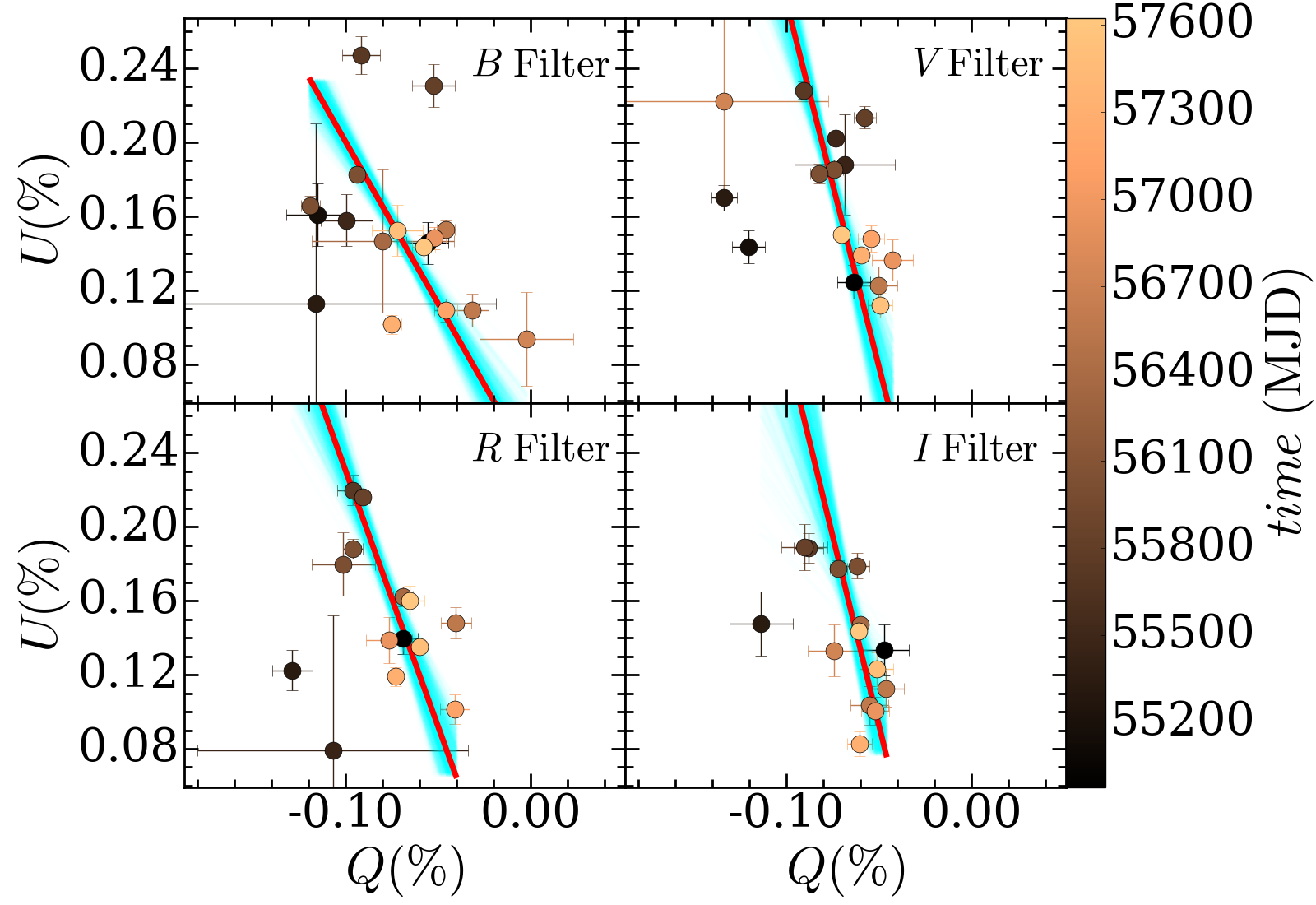}
    \caption{$Q-U$ diagram of polarization data of the fourth cycle of $\omega$ CMa. The date of observations are indicated in the legend on the right. The red solid lines are linear fits to the data. The cyan bands show the range of uncertainties in the fit. The data are binned in 100-day time intervals.}
    \label{fig:pol_qu}
\end{figure}



Figure~\ref{fig:pol_qu} shows the $Q-U$ diagram of the polarization data of C4. The original data have some individual points with significant variations and large error. Therefore, we binned the data in time intervals of 100 days. For all four bands, the measurements form a straight path in the $Q-U$ diagram as explained above. A simple linear least squares regression fit (solid red lines in Figure~\ref{fig:pol_qu}) indicates that the angle of this path is 117$^\circ \pm 7.6^\circ$, 102$^\circ \pm 5.8^\circ$, 110$^\circ \pm 4.6^\circ$, and 104$^\circ \pm 4.8^\circ$ for $B$, $V$, $R$, and $I$, respectively, which means that $\theta_{\rm int}$ should be half of this value. The errors were estimated using $\pm 1 \sigma$ uncertainty. These numbers are listed in Table~\ref{table:pol_angle} with the $QU$ superscript, to indicate that they were obtained using the $Q-U$ diagram method. The average value of the intrinsic position angle for the four filters is 54.2$^\circ \pm 2.9^\circ$, where the quoted uncertainty is the standard deviation of the mean.

The $Q-U$ method requires that the position angle of the disk remains relatively constant over time. We can estimate the validity of this assumption by measuring the correlation between the Stokes $Q$ and $U$. $\rho\approx -1$ as measured, e.g., by the method of Pearson correlation coefficient indicates a strong correlation. A weak correlation, i.e., $\rho\approx 0$, means the disk behavior is complex and the validity of the method is compromised. We obtained $\rho =$ -0.34, -0.51, -0.35, and -0.60 for $B$, $V$, $R$, and $I$ filters, respectively, which infer that the errors derived by the linear least squares regression fit are underestimated. However, our results suggest that the intrinsic polarization angle of the star might be close to 52$^{\circ}$ corresponding to the correlation coefficient closest to -1 ($I$-band).


\begin{table*}
    \begin{center}
        \caption{Different estimates of the IS polarization and the position angle of the intrinsic polarization. $QU$, LC, and field indicate the $Q-U$ diagram, the light curve, and the field star, respectively, as the methods that have been used to estimate the intrinsic polarization of $\omega$ CMa. The numbers in bold give the average value of the intrinsic position angle of four different filters measured by each particular method.}
\begin{tabular}{@{}ccccccc}
\hline
\hline
Method & Parameter & $B$ Filter & $V$ Filter & $R$ Filter & $I$ Filter & average \\
\hline
$QU$ &  $\theta_{\rm int}^{QU}$($^\circ$) & 58.4~$\pm$~3.8 & 51.2~$\pm$~2.9 & 55.2~$\pm$~2.3 & 52.0~$\pm$~2.4 & {\bf 54.2}~$\pm$~2.9 \\
\hline
 &  $P_{\rm IS}^{\rm LC}$(\%) & 0.12~$\pm$~0.01 & 0.15~$\pm$~0.01 & 0.11~$\pm$~0.01 & 0.11~$\pm$~0.01 & 0.12~$\pm$~0.02 \\
LC &  $\theta_{\rm IS}^{\rm LC}$($^\circ$) & 56.4~$\pm$~1.5 & 53.5~$\pm$~1.8 & 56.0~$\pm$~2.1 & 56.6~$\pm$~1.9 & 55.6~$\pm$~1.2 \\
 &  $\bar{\theta}_{\rm int}^{\rm LC}$($^\circ$) & 56.2~$\pm$~10.9 & 62.9~$\pm$~9.8 & 63.1~$\pm$~2.8 & 58.8~$\pm$~9.3 & {\bf 60.2}~$\pm$~2.9 \\
\hline
 &  $P_{\rm IS}^{\rm field}$(\%) & 0.13~$\pm$~0.02 & 0.12~$\pm$~0.03 & 0.11~$\pm$~0.01 & 0.08~$\pm$~0.01 & 0.11~$\pm$~0.02 \\
Field &  $\theta_{\rm IS}^{\rm field}$($^\circ$) & 59.7~$\pm$~4.6 & 53.5~$\pm$~6.6 & 57.9~$\pm$~2.9 & 60.8~$\pm$~1.6 & 58.0~$\pm$~2.8 \\
 &  $\bar{\theta}_{\rm int}^{\rm field}$($^\circ$) & 54.5~$\pm$~11.3 & 62.7~$\pm$~10.7 & 58.2~$\pm$~8.0 & 55.0~$\pm$~11.6 & {\bf 57.6}~$\pm$~3.3 \\
\hline
\end{tabular}
\label{table:pol_angle}
\end{center}
\end{table*}


\subsubsection{Light curve method}
\label{subsubsect:lc}

The second method uses the light curve itself. The results of \citetalias{ghoreyshi2018} indicate that at the end of C4, the $V$-band excess is very small. According to the best \citetalias{ghoreyshi2018} model fits, the inner disk at that phase is very tenuous (Figure~\ref{fig:summary}, see also second panel of Figure~11 of \citetalias{ghoreyshi2018}). Therefore, if one assumes that $P_{\rm int}$ at that phase is very small, the observed polarization should be very close to $P_{\rm IS}$. According to the relatively nearby distance of $\omega$ CMa, small values for $P_{\rm IS}$ are expected \citep[e.g.,][]{serkowski1975, yudin2001}.

\begin{figure}
    \centering
    \includegraphics[width=\columnwidth]{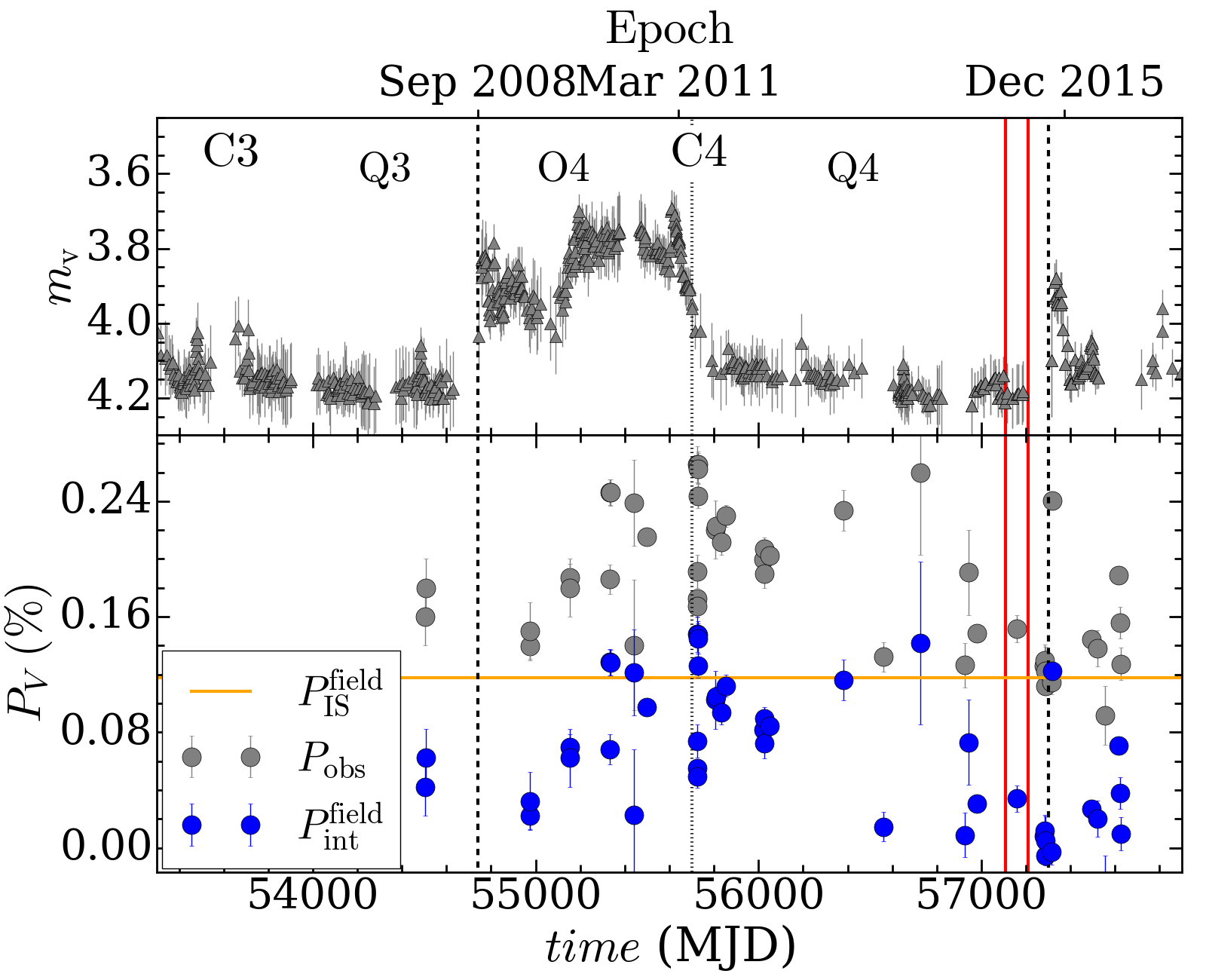}
    \caption{{\it Top}: The $V$-band observed photometric data for $\omega$ CMa. Two vertical red lines indicate the epoch during which the star was assumed to be diskless. {\it Bottom}: The observed (grey circles) and intrinsic polarization (blue circles) of the fourth cycle of $\omega$ CMa in the $V$ filter, measured by the field star method. The estimate of $P_{\rm IS}^{\rm field}$ is shown by the horizontal orange line. C$i$, O$i$ and Q$i$ stand for cycle, outburst and quiescence phases, respectively, where $i$ is the cycle number. The vertical dashed lines display the boundaries between the cycles. The vertical dotted lines show the boundaries between outburst and quiescence phases.}
    \label{fig:pol_v_corrected}
\end{figure}

Figure~\ref{fig:pol_v_corrected} presents the photometric and polarimetric data of $\omega$ CMa in the $V$ band. The observed polarization is shown as grey circles. Two vertical red lines in Figure~\ref{fig:pol_v_corrected} indicate the boundaries of the phase assumed for the star to be almost diskless. An average of the data at this phase (a single point for the $B$, $V$, and $R$ filters, and a few points for the $I$ filter; see Figure~\ref{fig:pol_bvri_corrected}), gives us $Q_{\rm IS}^{\rm LC}$ and $U_{\rm IS}^{\rm LC}$, which in turn provide the values for $P_{\rm IS}^{\rm LC}$ and $\theta_{\rm IS}^{\rm LC}$, listed in Table~\ref{table:pol_angle} for each filter. Here the superscript ${\rm LC}$ indicates that the estimates were made using the light curve itself. Using Eqs.~\ref{eq:p} to~\ref{eq:u_obs} and the estimated value of interstellar polarization, the intrinsic polarization of $\omega$ CMa, $P_{\rm int}$ and $\theta_{\rm int}$, was calculated. The average value of $\bar{\theta}_{\rm int}^{\rm LC}$ is 60.2$^\circ \pm 2.9^\circ$, in good agreement with the value estimated using the $Q-U$ method. We defer for later a discussion on the intrinsic polarization levels.

%

\subsubsection{Field star method}
\label{subsubsect:fs}

Finally, we estimated the $P_{\rm IS}$ using the field star method, by which one or more stars that are physically close to the target star and that are known to have no intrinsic polarization are used as a proxy of the interstellar polarization.
By the IAGPOL, we observed HD56876, a B5Vn star \citep{houk1982} with $m_{\rm v} \approx$ 6.4, a Gaia distance of 282$^{+4}_{-3}$pc and an angular distance of 0.67$^\circ$ from $\omega$ CMa. The distance of $\omega$ CMa inferred from the parallax measured by Gaia is 280$^{+13}_{-11}$~pc \citep{gaia2016, gaia2020}.

The results of this method are listed in Table~\ref{table:pol_angle} with the ``field'' superscript. The agreement with the previous method based on the light curve is quite good, as both results have very similar values for $P_{\rm IS}$ and $\theta_{\rm IS}$. Furthermore, the three estimates of $\theta_{\rm int}$ (shown by bold numbers in Table~\ref{table:pol_angle}) also agree. By modeling the Br-$\gamma$ interferometric data for $\omega$ CMa, \cite{stefl2011} showed that the position angle of the major axis of the star's disk is -29$^{\circ}$ (see the lower left panel of Figure~1 in their paper). Since there is a 90$^{\circ}$ difference between the elongation of the minor and major axes, this means the position angle for the minor axis of the disk is 61$^{\circ}$ which is in agreement with the results presented here.

The estimated $P_{\rm int}$ level from the field star method is illustrated with blue circles in the bottom panel of Figure~\ref{fig:pol_v_corrected}. A positive correlation between the polarization level and the brightness of the star can be seen. It appears that the polarization level follows the variation of the $V$-band photometric data with a lag, for instance, in the dissipation phase the drop of polarization is slower than the drop in brightness. This agrees with the theoretical studies of \cite{haubois2014}. This behavior is easily explained when we consider that the $V$-band polarization probes a slightly larger volume of the disk than the $V$-band (see Figure~1 of \citealt{carciofi2011}) and, therefore, the timescales for viscous dissipation will be longer. There is also some intrinsic scatter in the data that is likely the result of disk variability. As shown by \cite{haubois2014}, the polarization level responds quickly to changes in the $\dot J$. The flickers seen in the $V$-band light curve of $\omega$ CMa should, therefore, have a polarimetric counterpart. Indeed, some of the polarization variability seems to be directly related to flickering events (e.g., epochs 56300 to 56800).

Below we use the values of the field star method, because the light curve method can have systematic errors if the $P_{\rm int}$ at the epochs chosen (bracketed by the red lines in Figure~\ref{fig:pol_v_corrected}) is non-zero. In fact, our model calculations (see Figure~\ref{fig:polarimetry_model} below) indicate that this may be the case. The same can be said about the field star method, of course, as the field star may not probe the same $P_{\rm IS}$ as the target star. Our choice for the field star method is thus based solely on the fact that one method (the light curve one) very likely suffers from systematic errors while for the other method this is unknown.

\subsubsection{Model-data comparison}
\label{subsubsect:md}

\begin{figure}
\begin{minipage}{\linewidth}
\centering
\subfloat[]{\includegraphics[width=1.0\linewidth]{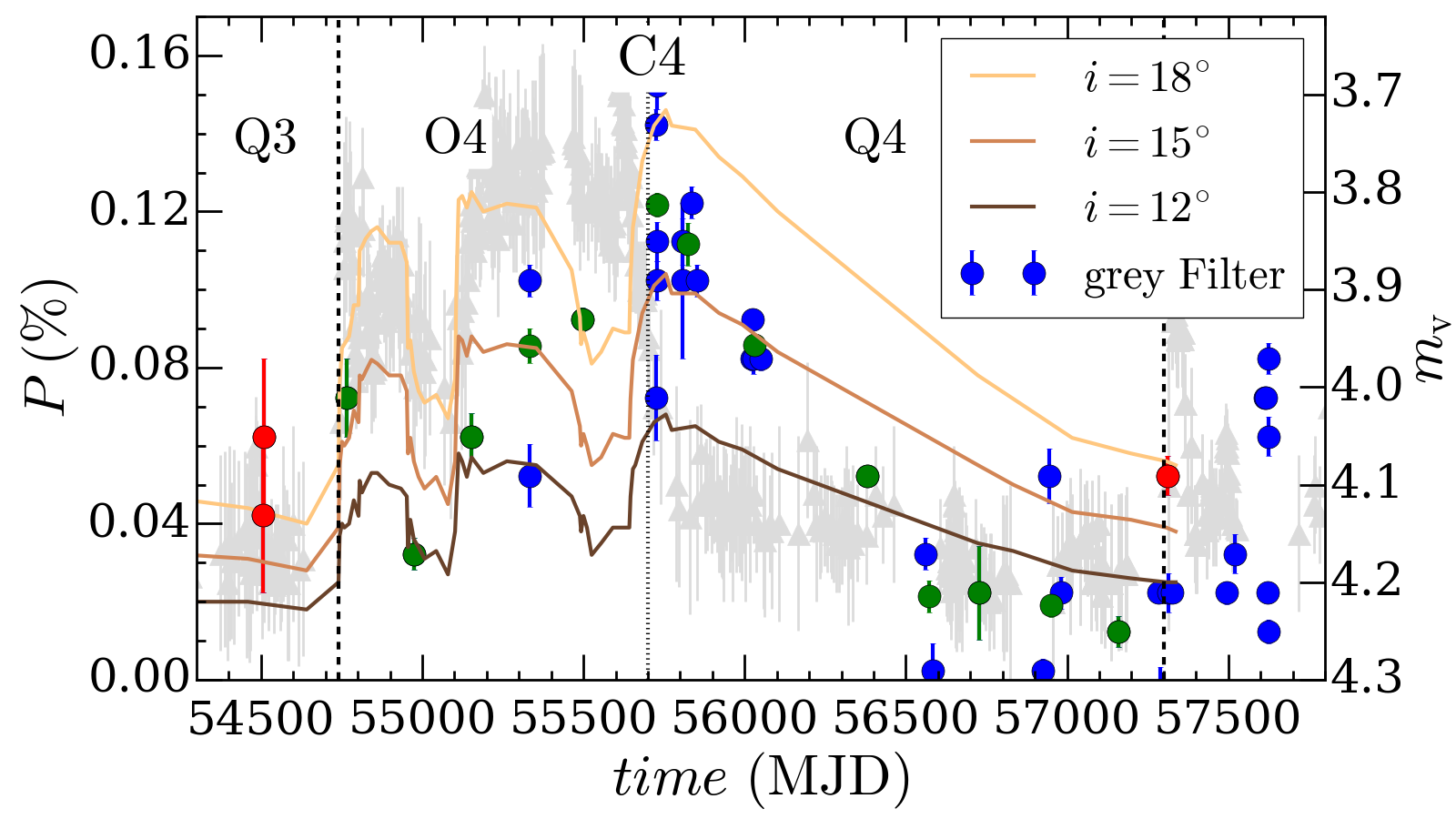}}
\end{minipage}%
\begin{minipage}{\linewidth}
\centering
\subfloat[]{\includegraphics[width=1.0\linewidth]{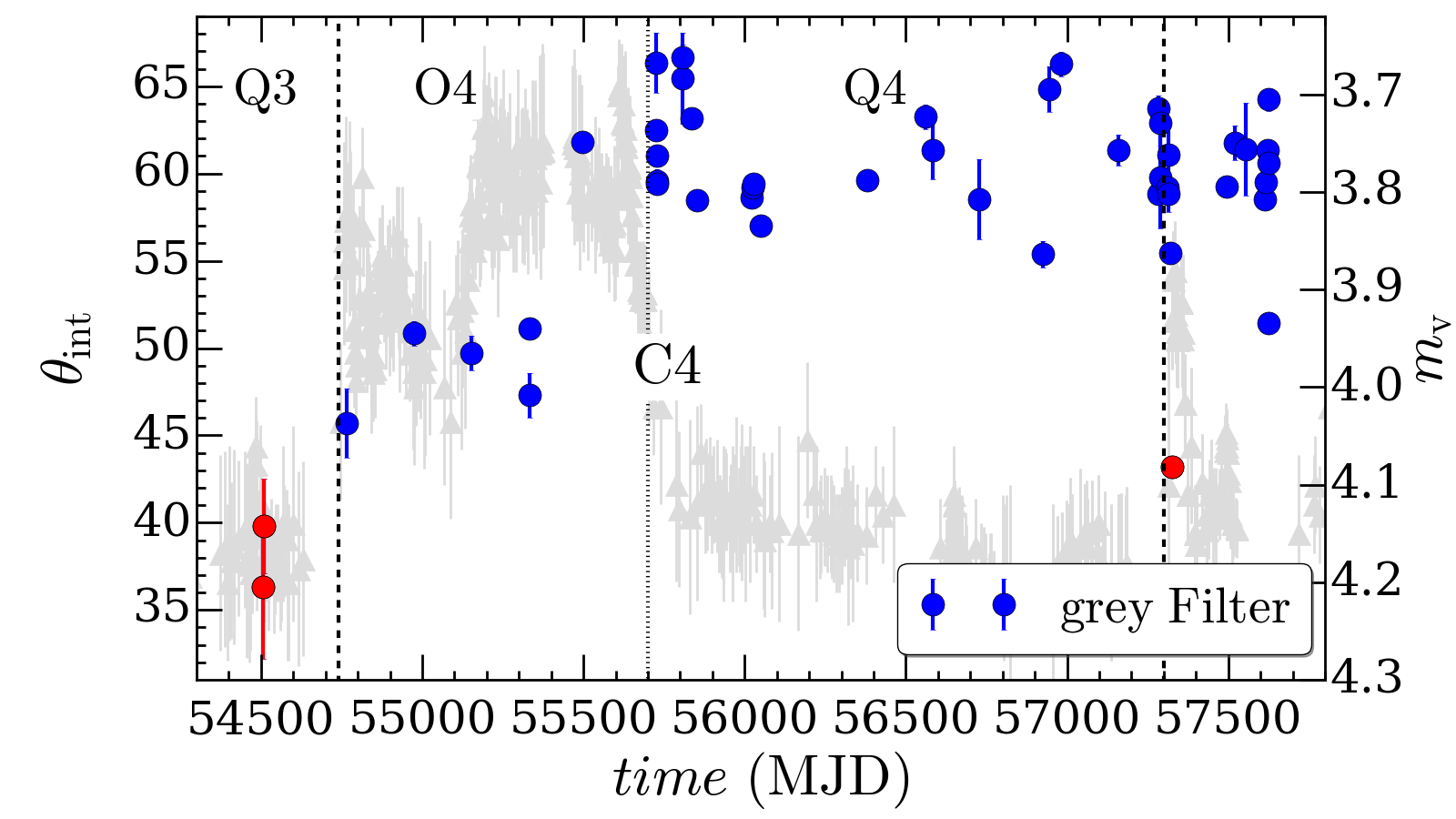}}
\end{minipage}
\caption{(a) The averaged polarimetric data of $\omega$ CMa vs.\ the model for the last cycle. The data are presented with blue circles. The colored lines represent the synthetic polarization, with the colors representing different inclinations as indicated in the legend. The best fit is for $i=15^{\circ}$ with $\chi^{2}_{\rm red}=50$ compared with 107 and 215, for $i=12^{\circ}$ and $18^{\circ}$, respectively. For calculating $\chi^{2}_{\rm red}$ the data were binned (green circles) in time intervals of 100 days to combine the observed points in the vicinity of epochs for which the model was computed. (b) The averaged polarization position angle of $\omega$ CMa. In all panels, the red points indicate data taken in epochs with a clear photometric flicker. The $V$-band photometric data are shown with grey triangles in the background. C$i$, O$i$ and Q$i$ stand for cycle, outburst and quiescence phases, respectively, where $i$ is the cycle number. The vertical dashed lines display the boundaries between the cycles. The vertical dotted lines show the boundaries between outburst and quiescence phases.}
\label{fig:polarimetry_model}
\end{figure}

The top panel of Figure~\ref{fig:polarimetry_model} displays the \citetalias{ghoreyshi2018} model for three different inclination angles. To increase the signal-to-noise (S/N) of the data, we use an inverse-error-weighted average of all filters (i.e. grey filter) available for each epoch. The model was also averaged in the same way, to ensure a proper comparison with the data. The agreement between the \citetalias{ghoreyshi2018} best fit model (for which $i=15^{\circ}$ and reduced $\chi^{2}$, $\chi^{2}_{\rm red}=50$) and the data from C4 is reasonable. Recall that this model was developed based solely on the $V$-band photometry, so the broad agreement for the polarization level is encouraging. The model seems to reproduce the variations due to the short (partial) formations and dissipations during the main formation phase (O4). However, it is apparent that the model cannot reproduce the rate of polarimetric variations during Q4, as $P_{\rm int}$ drops faster than the model does. The implications of this result will be further discussed in Section \ref{subsect:spectroscopy}.


In the bottom panel of Figure~\ref{fig:polarimetry_model} we display $\theta_{\rm int}$, averaged for all filters (i.e., grey filter) in the same way as was done for $P$. Interestingly, there is clear evidence for a trend in $\theta_{\rm int}$, from about 40$^{\circ}$ at the beginning of O4 to 60$^{\circ}$ towards the end of this phase. After that the position angle remains constant, with the exception of one point near MJD = 57300. This last point is marked in red, as well as two points at MJD = 54500, to indicate that they coincide precisely with a photometric flicker in the light curve. The behavior of $\theta_{\rm int}$ might be an indication of matter being injected in the disk outside of the equatorial plane. Evidence for this comes from the fact that whenever matter is fed into a low-density disk (red points and the beginning of O4) the angle is different than when a fully formed disk is present (end of O4) or no matter is being ejected (Q4). In this scenario, the position angle of the disk would be about 60$^{\circ}$, while the position angle of the injected matter would be roughly 20$^{\circ}$ different. It should be emphasized that the above trend for $\theta_{\rm int}$ can only be seen in the averaged data, and is not discernible in the data for each filter (Figure~\ref{fig:pol_bvri_corrected}).

Finally, similar to the polarization level, the polarization angle has a large scatter, much larger than the observational errors. The scatter in the data
may also be related to the disk feeding process: depending on how matter is ejected from the star, an axial asymmetry may develop in the inner disk, which can cause large variations in the polarization position angle. For instance, \cite{carciofi2007} detected changes in $P_{\rm int}$ of up to 8$^\circ$ in less than one hour for the star Achernar, following putative mass loss events. Similar variations in $\theta_{\rm int}$ have been observed for other Be stars, and some of these could be related to specific outburst events \citep[see discussion in section 4 of][]{draper2014}.

The similarity of the results for each method that have been discussed in Sections \ref{subsubsect:qu} to \ref{subsubsect:fs} suggests that the results presented here would not be altered with minor changes in $P_{\rm IS}$.




\subsection{Magnitudes and colors}
\label{subsect:mag_color}

In this section our comparison of the model with photometric data in wavelengths other than the visible is presented.

\begin{figure}
    \centering
    {\includegraphics[width=1.0 \linewidth]{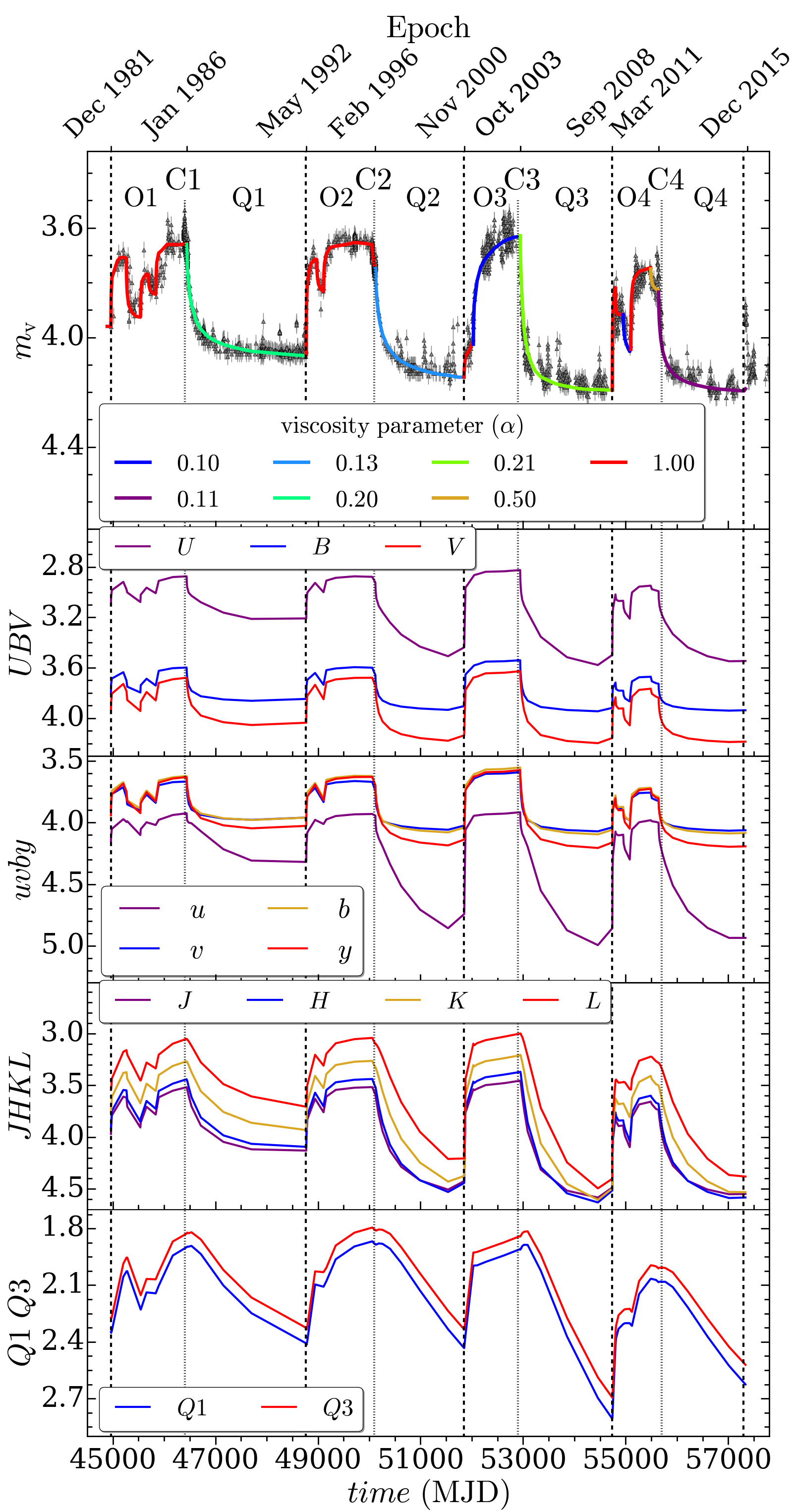}}
    \caption{Synthetic light curves of $\omega$ CMa at different bands from the UV to the far IR. The upper plot shows the fit of the VDD model to the $V$-band photometric data (see Figure~11 of \citetalias{ghoreyshi2018} for more details). The other plots display the model light curves for different band-passes as indicated. The vertical dashed lines display the boundaries between the cycles. The vertical dotted lines show the boundaries between outburst and quiescence phases.}
    \label{fig:mag_color}
\end{figure}

Figure~\ref{fig:mag_color} displays the synthetic light curves of $\omega$ CMa. The top panel of Figure~\ref{fig:mag_color} illustrates the $V$-band modeling presented in \citetalias{ghoreyshi2018}. The second panel shows the synthetic {\it UBV} bands together. Interestingly, the $U$-band light curve displays the largest variations, which is expected as this band should be more sensitive to the inner disk conditions, where the density varies widely (see Figure~11 in \citetalias{ghoreyshi2018}). The model predicts a complex behavior. In general, we see that the longer the wavelength, the slower the rate of magnitude variations in the model light curve. This is explained by the fact that larger disk volumes (from where the long wavelength continuum fluxes originate; see Figure~7 of \citealt{rivinius2013a}) respond more slowly to variations in the inner disk. One interesting feature of the models is that each subsequent dissipation reached a lower flux level when compared to the previous. This reflects the finding in \citetalias{ghoreyshi2018} that in $\omega$ CMa a true quiescence value is never realized, but rather the star transitions between high and low mass loss rate states.


\begin{figure*}
    \begin{minipage}{0.5\linewidth}
        \centering
        \subfloat[]{\includegraphics[width=1.0\linewidth]{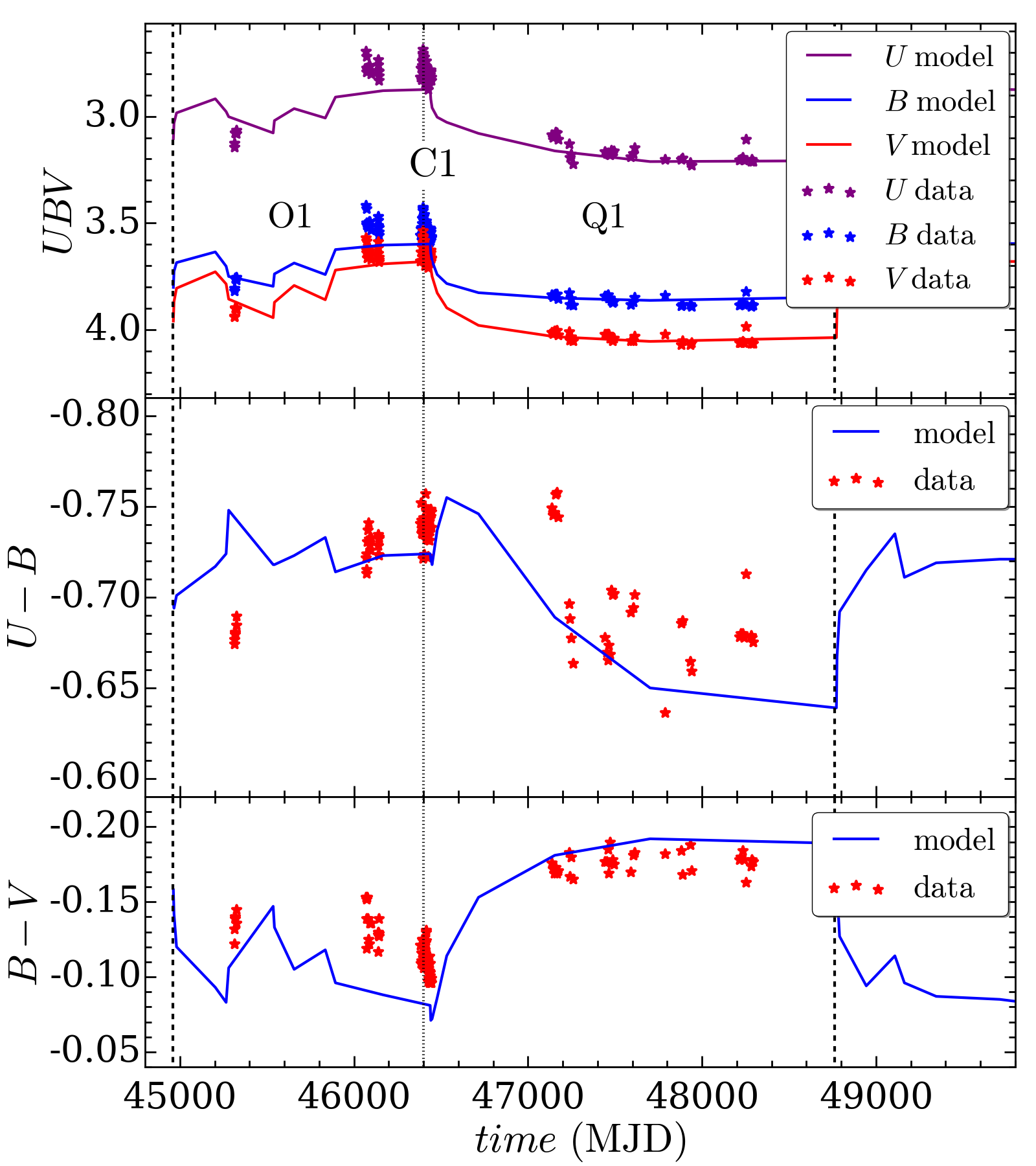}}
    \end{minipage}%
    \begin{minipage}{0.5\linewidth}
        \centering
        \subfloat[]{\includegraphics[width=1.0\linewidth]{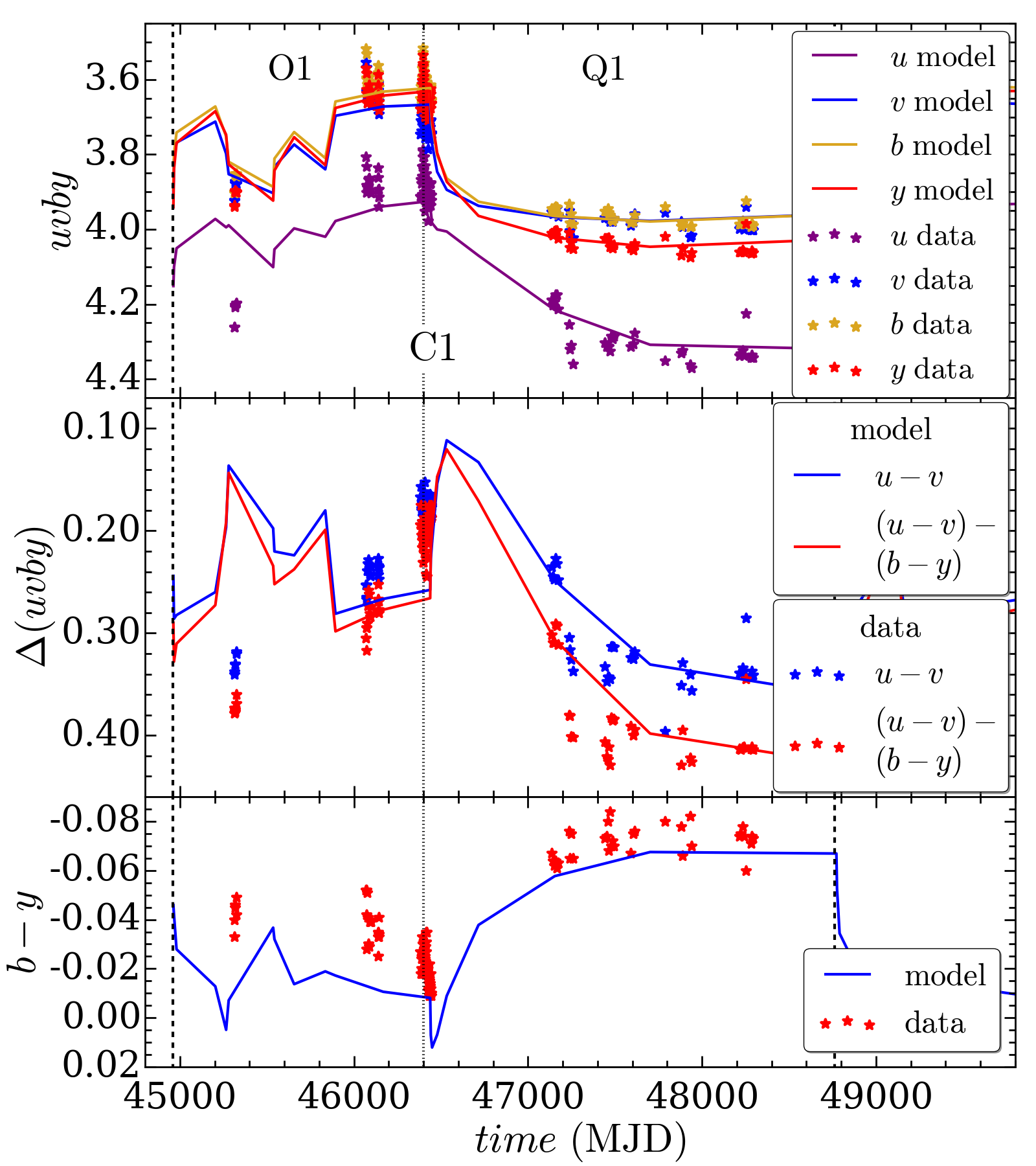}}
    \end{minipage}
    \caption{Comparison between the observed magnitudes and color-indices (stars) and the VDD model (lines) (a) {\it UBV} filters (b) Str\"{o}mgren {\it uvby} filters, both from LTPV. C1, O1 and Q1 stand for the first cycle, outburst and quiescence phases, respectively. The models shown here are limited only to the first cycle because we do not have data for the rest. The vertical dashed lines display the boundaries between the cycles. The vertical dotted lines show the boundaries between outburst and quiescence phases.}
    \label{fig:ltpv_time}
\end{figure*}

The left panel of Figure~\ref{fig:ltpv_time} shows the comparison between the observed {\it UBV} data and color-indices with the model. The model fits the data generally well. There are a few outliers, which are likely explained by the fact that we did not model the short term flickering events (i.e., events shorter than two months, see \citetalias{ghoreyshi2018}) of the light curve as mentioned previously. A systematic mismatch, however, is seen in the color indices, most notably at $B-V$.

The Str\"{o}mgren LTPV data ($uvby$-bands) are demonstrated in the right panel of Figure~\ref{fig:ltpv_time}. The general behavior is similar to the {\it UBV} data. The general shape of the curve, as well as the colors, are however quite well reproduced by the model.


The rough agreement between the VDD model and the $uvby$-bands data is a significant result, as these band-passes probe slightly different regions of the disk. In general the flux level is related to the disk density, while the colors probe the density gradients. Our results indicate that the density scale of the inner disk is well-reproduced by the model, but the mismatch in the colors may point to inaccuracies in the density gradient. This is not surprising, given the simplistic nature of our assumptions for the disk inner boundary conditions (see \citetalias{ghoreyshi2018} and \citealt{rimulo2018}, for more details).

\begin{figure}
    \begin{minipage}{\linewidth}
        \centering
        {\includegraphics[width=\columnwidth]{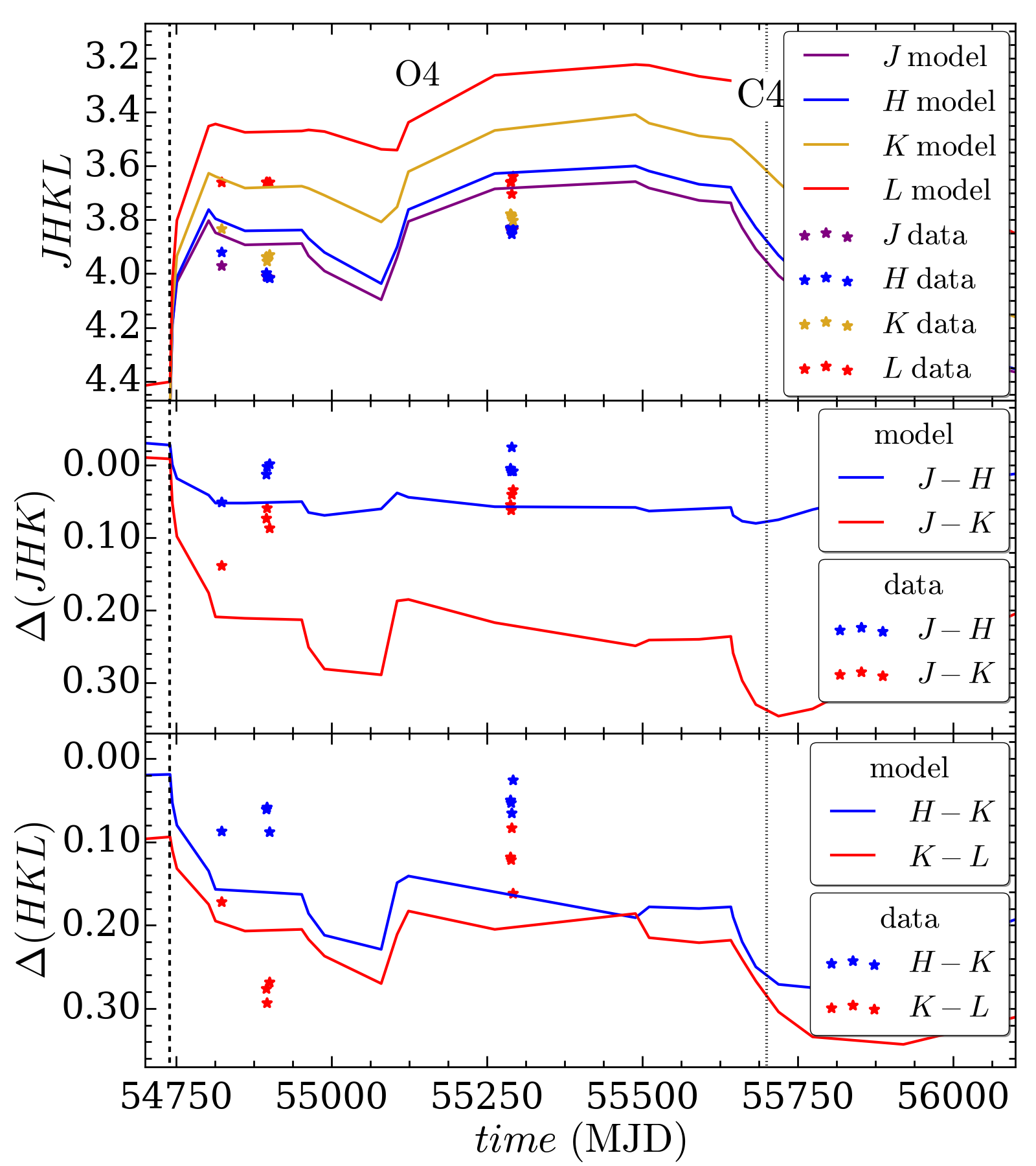}}
    \end{minipage}
    \caption{Comparison between magnitudes and color indices of the VDD model and observed data in the {\it JHKL} filters. The VDD models are demonstrated with solid lines and the data are shown with colored stars as indicated in the legend of the plot. C4 and O4 stand for the fourth cycle and outburst phase, respectively. The models presented here are limited to this cycle for which we have data. The vertical dashed lines display the boundaries between the cycles. The vertical dotted lines show the boundaries between outburst and quiescence phases.}
    \label{fig:jhkl_time}
\end{figure}

Figure~\ref{fig:jhkl_time} shows the comparison between the observed {\it JHKL} magnitudes, colors and the model. The top panel reveals that the model is consistently brighter than the data. This discrepancy is maximum for the $L$ band and minimum for the $J$ band. Thus, the longer the wavelength, the larger the discrepancy. The middle and bottom panels display that the color indices are also, in general, systematically shifted with respect to the models. It should be noted that it is unlikely that the discrepancies seen for the {\it JHKL} bands are due to uncertainties in the inclination angle. For instance, changing the inclination angle to 18$^\circ$ in the models would cause a magnitude increment of only $\approx$ 0.02.


\begin{figure}
    \begin{minipage}{\linewidth}
        \centering
        {\includegraphics[width=\columnwidth]{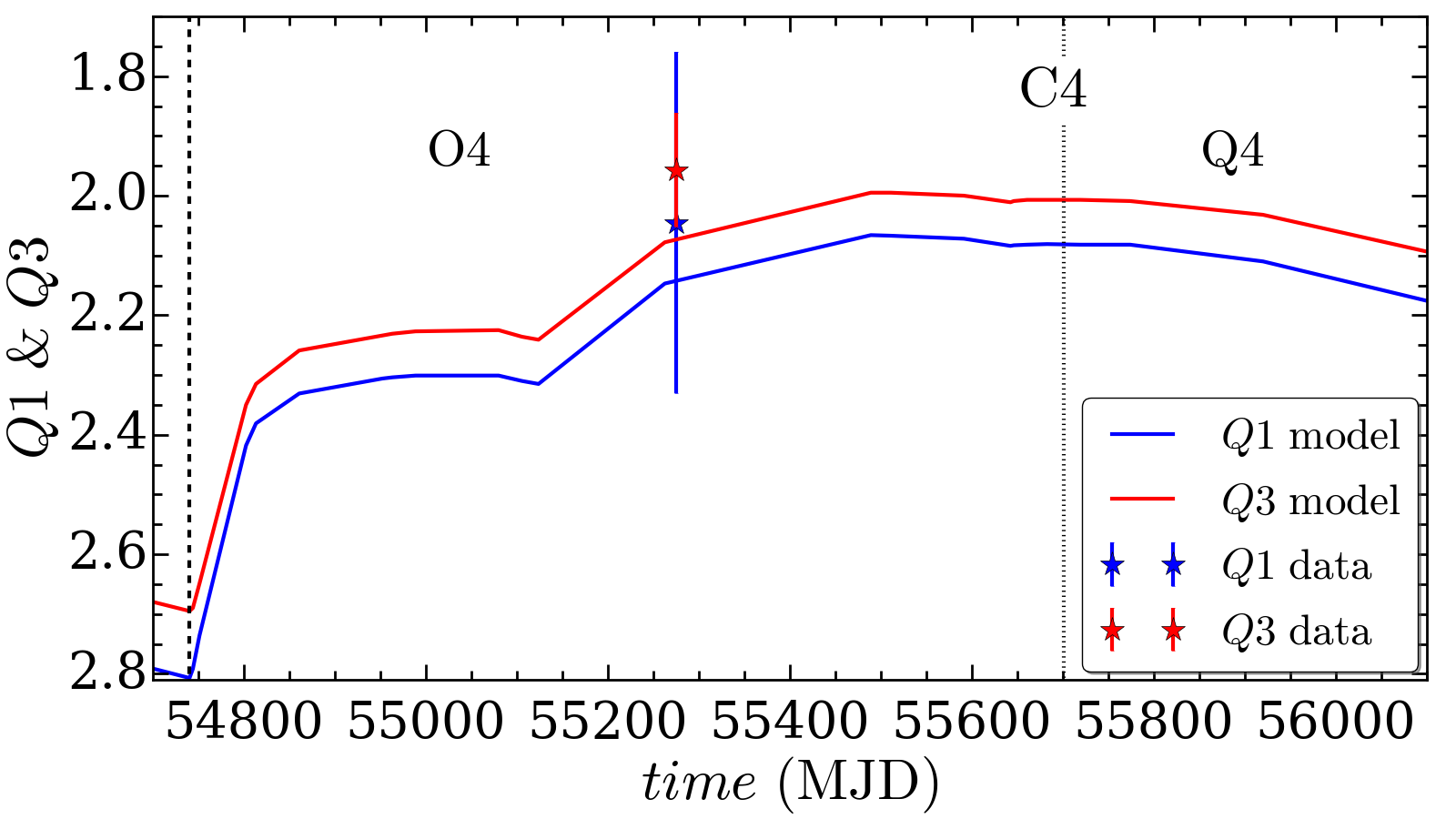}}
    \end{minipage}
    \caption{Comparison between magnitudes of the VDD model and the observed data in the $Q1$ and $Q3$ filters. The VDD models are demonstrated with solid lines and the data are shown with colored stars as indicated in the legend of the plot. C4, O4 and Q4 stand for the fourth cycle, outburst and quiescence phases, respectively. The models presented here are limited to this cycle for which we have data. The vertical dashed lines display the boundaries between the cycles. The vertical dotted lines show the boundaries between outburst and quiescence phases.}
    \label{fig:q_time}
\end{figure}


Although the scarcity of IR data makes a firm conclusion difficult, it still appears that the {\it JHKL} color indices are more or less well reproduced by the VDD model, while the actual magnitudes are not. The model is always systematically brighter in the {\it JHKL} bands than the observations, which means that the model may be too dense in the outer disk regions. This might suggest a larger $\alpha$ in the outer disk that would drain it faster, making it less dense. This is discussed in Section~\ref{sect:alternative}.

Figure~\ref{fig:q_time} displays the model and data in the VISIR $Q1$ ($\sim$16.7--18.0 $\mu$m) and $Q3$ ($\sim$19.1--20.0 $\mu$m) bands. Unlike the {\it JHKL} magnitudes, the data and models agree within the errors. Because the observational errors are large and only two points of the $Q1$ and $Q3$ data in total are available, these data do not allow us to confirm or disprove the tendencies seen for the {\it JHKL} bands.


\subsection{Spectroscopy}
\label{subsect:spectroscopy}

\begin{figure*}
    \begin{minipage}{0.5\linewidth}
        \centering
        \subfloat[H$\alpha$]{\includegraphics[width=1.0\linewidth]{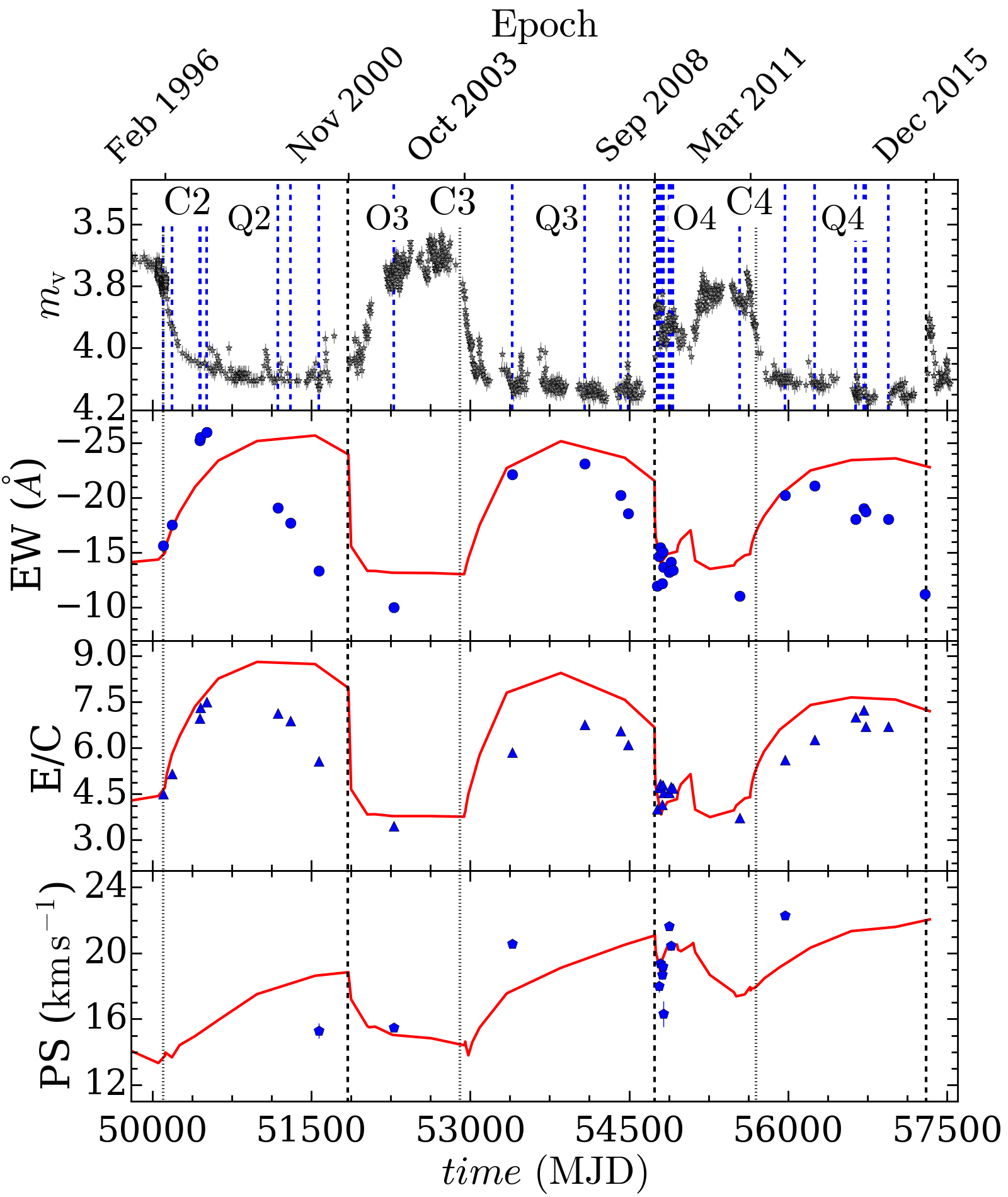}}
    \end{minipage}%
    \begin{minipage}{0.5\linewidth}
        \centering
        \subfloat[H$\beta$]{\includegraphics[width=1.0\linewidth]{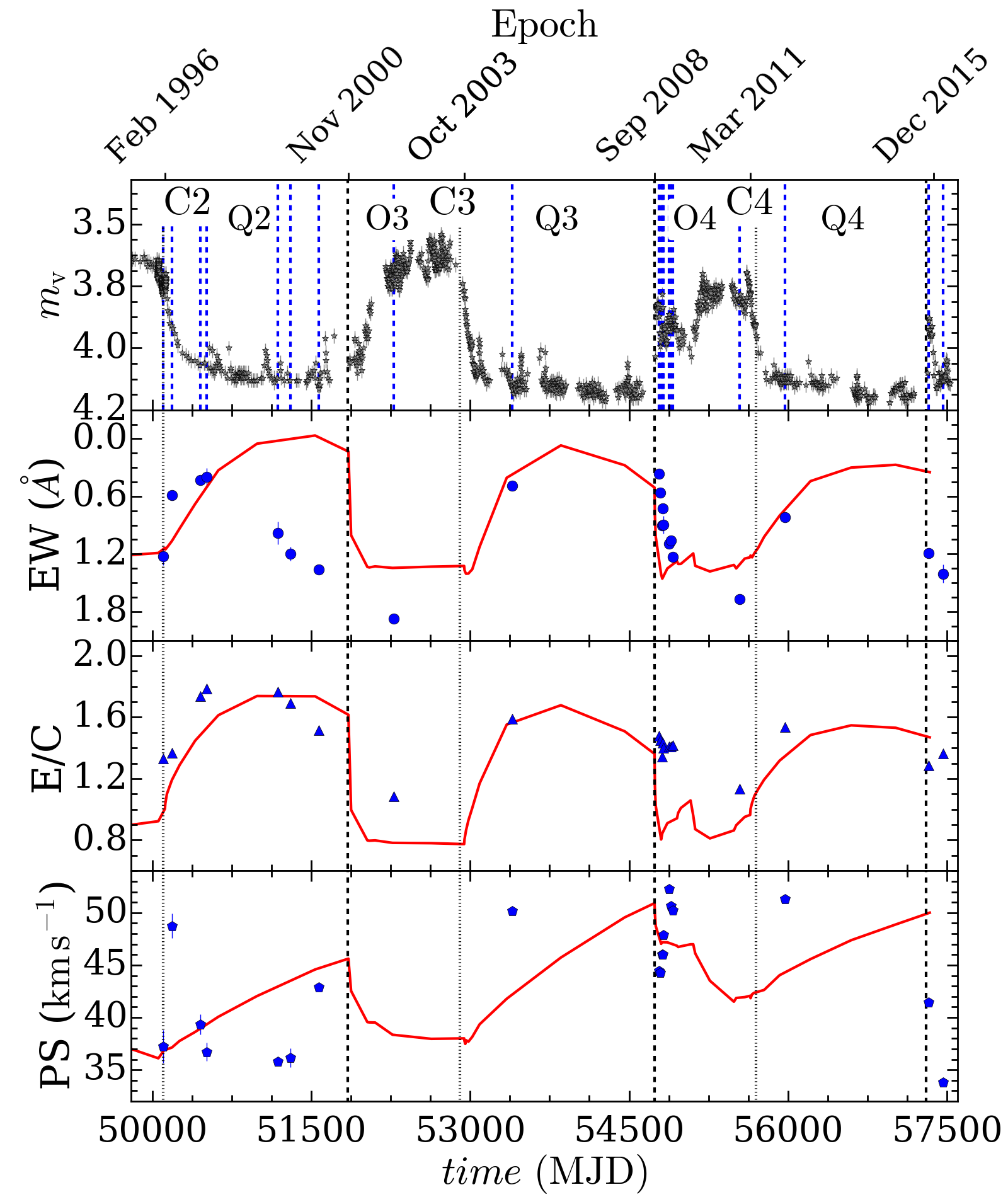}}
    \end{minipage}
    \caption{Comparison of the models (lines) and the (a) H$\alpha$ and (b) H$\beta$ EW, E/C, and PS. The top plot in each panel displays the $V$-band light curve with vertical blue dashed lines marking the date the spectrum was taken. Observed EW, E/C, and PS are shown with blue circles, triangles, and pentagons, respectively, in the second to the fourth panels. C$i$, O$i$ and Q$i$ stand for cycle, outburst and quiescence phases, respectively, where $i$ is the cycle number. Models for the first cycle are not presented due to lack of data. The vertical dashed lines display the boundaries between the cycles. The vertical dotted lines show the boundaries between outburst and quiescence phases.}
    \label{fig:spectroscopy_model_balmer}
\end{figure*}

We have a comprehensive dataset of spectra for $\omega$ CMa covering the cycles 3 and 4 (Figure~\ref{fig:data_dist}c), as well as Q2. As examples, we show the comparison between the VDD model and observed equivalent width (EW), E/C, and peak separation (PS) for H$\alpha$ and H$\beta$ in Figure~\ref{fig:spectroscopy_model_balmer}. The results for H$\gamma$ and H$\delta$ are qualitatively similar, and are shown in Appendix~\ref{ap:spectro}.

These spectra have good S/N and medium-to-high resolution. Thus, the uncertainties of the quantities are small. The uncertainties for the line profile E/C and EW are dominated by their continuum determination, while for the PS the resolution is the limiting factor.


It is important to emphasize that we did not adjust the model to obtain an optimum fit. We used the same model and scenario described in \citetalias{ghoreyshi2018} to calculate these profiles, in order to evalutate how this model performs when compared to multi-technique data.

The second and third panels of each plot in Figure~\ref{fig:spectroscopy_model_balmer} display the EW and E/C of the line. The EW and the E/C ratio of a line reveal a complex interplay between the line emission and that of the adjacent continuum. The H$\alpha$ line emission comes from a large volume of the disk, which responds very slowly to changes in the disk feeding rate. Conversely, the adjacent continuum responds very quickly to these changes. Therefore, when the continuum emission rises (e.g., during an outburst), the EW initially drops in magnitude and the E/C ratio falls, as well. However, when the continuum emission drops (e.g., during quiescence), the EW will increase in magnitude and the E/C ratio will increase. These effects are more moderate for H$\beta$ since at 4861\,\AA~the adjacent continuum displays a much smaller range of magnitude variation along a given cycle than at 6562\,\AA~(Figure~\ref{fig:mag_color}).

Figure~\ref{fig:spectroscopy_model_balmer} can be interpreted with the above scenarios in mind. In all quiescence phases, the EW increases in magnitude (becoming more negative) and the E/C increases, as a result of the quick suppression of the inner disk, that causes the emission in the adjacent continuum to drop quickly. This initial dissipation of the inner disk does not affect the line emission. Only much later in the dissipation when the entire disk empties, the line emission drops. Then, the EW decreases in magnitude and E/C also drops. At outburst, the converse happens: the inner disk fills up quickly, giving rise to a sudden increase in the continuum out to the IR. As a result, the EW decreases in magnitude and E/C decreases (recall the apparent contradiction mentioned at the end of Section~\ref{sect:omecma_obs} and also, see Figure~\ref{fig:feros}).

The PS was computed by fitting a Gaussian curve to each emission peak, in order to determine its height compared to the adjacent continuum (which was normalized to one), as well as its velocity. The low inclination angle of $\omega$ CMa causes an almost single-peak profile in H$\alpha$ whose flux comes mainly from the larger part of the disk (in comparison to the other hydrogen lines) where the Keplerian velocities are lower. Also, some of the data are of low resolution, which makes our analysis difficult. For this reason we show the PS in Figure~\ref{fig:spectroscopy_model_balmer} only for the spectra with a clear double-peaked structure. Comparison with the model reveals that, similarly to what was seen for the EW and E/C, there is a better agreement during the outburst phases than during dissipation.

In general, the results for H$\beta$ are similar to those for H$\alpha$. The EW curve is qualitatively reproduced, but a quantitative comparison fails mainly during the quiescence phases. Of particular significance is the close match between the data and the model for the fast decline in EW of O4. The E/C is also well reproduced. It is important to recall that, since the H$\beta$ opacity is smaller than H$\alpha$, the formation volume of this line is smaller \citep{carciofi2011}. This can be seen by the PS values, which lie $\approx$ 40 $\rm{km\,s^{-1}}$ for H$\beta$, while for H$\alpha$ they are, in general, smaller than 20 $\rm{km\,s^{-1}}$. The larger PS indicate that H$\beta$ is indeed formed closer to the star, where the rotational velocities are larger. The fact that the model reproduces this behavior is a significant result.

The model can reproduce these variations qualitatively, but not quantitatively. After the quick increase in magnitude of the EW at the onset of dissipation, the observed EW decreases in magnitude at a much faster rate than the model does (the same is observed with the E/C). Since the predictions from \citetalias{ghoreyshi2018} fit the visual and infrared band light curves well, the problem likely lies in the outer disk. It appears that while \citetalias{ghoreyshi2018}'s model predicts the correct rate of density variation in the inner disk, the corresponding rates in the outer part are too slow. In other words, the outer disk is not being drained of material fast enough.  
Further support to this comes from polarimetry. Recall that the observed rate of polarimetric variation is larger than the model calculations, also indicating faster emptying than in the model.

In the following, we provide some tentative explanation for this mismatch between the model and the data. 
One way to achieve faster dissipation rates in the outer disk is to have larger values of the viscosity; this could happen either because the temperature rises with radius (which is not physically justified) or the $\alpha$ parameter increases with distance from the star. Therefore, this might be the first hint of a radially varying $\alpha$ in a Be star. Another possibility is to consider that there is an unknown binary companion truncating the disk at radii smaller than the 1000$R_{\rm eq}$ assumed here (note that this was an arbitrary assumption for the disk size in \citetalias{ghoreyshi2018}). If this were the case, the mass reservoir of the outer disk would be smaller and the whole disk would dissipate faster, as suggested by the observations. Finally, a third possible explanation for the mismatch is radiative ablation. In the absence of active feeding, ablation could act in addition to viscosity to dissipate disk material. However, the results of \citet[][and subsequent papers]{kee2016a} indicate that ablation is more efficient at clearing the inner disk rather than the outer. Therefore, ablation, if included in our models, would likely make the mismatch between the rates of the dissipation of the inner and outer disk worse. In the next section, we investigate the first two possibilities, namely larger $\alpha$ in the outer disk and binary truncation. Also, we discuss the influence of ablation in more details.


\section{Testing alternative models}
\label{sect:alternative}

So far, we showed that the model presented in \citetalias{ghoreyshi2018} is unable to reproduce some of the characteristics of the observed spectra, multi-wavelength photometry and optical polarimetry. These discrepancies seem to indicate that the models predict an outer disk that is too massive and a rate of dissipation during quiescence phases that is too slow. Thus, we need to adjust our models so that the disk dissipation rate at larger radii is larger. We investigate two possible solutions, namely: 1) larger values of the $\alpha$ parameter, and 2) disk truncation by a binary companion.

In the first test, we compare the new models based on larger values of $\alpha$ with the polarimetric data and H$\alpha$ EW during the quiescence phase of the fourth cycle, C4, to probe the effectiveness of the method for two datasets originating from regions within the disk at greater radii from the central star. (We note that C4 is the only cycle for which we have simultaneous polarimetric and spectroscopic data.) To further support our findings, we verify this approach by using it to model also the H$\alpha$ EW during the second quiescence (Q2) since the EW data for Q2 show a clearer pattern.

\begin{figure}
    \begin{minipage}{\linewidth}
        \centering
        {\includegraphics[width=\columnwidth]{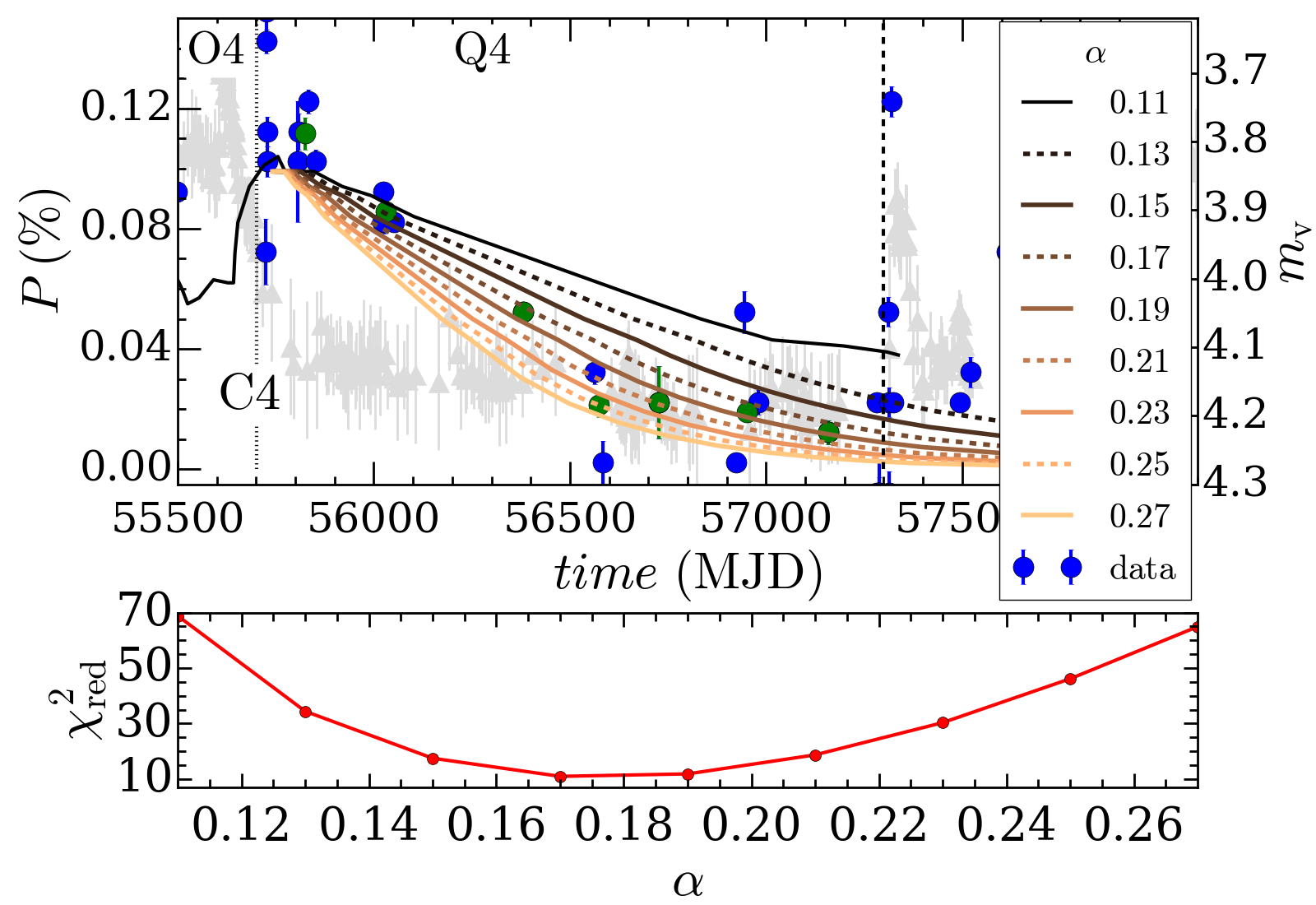}}
    \end{minipage}
    \caption{
    Top panel shows comparison between the models with various $\alpha$ values and observed polarimetric data. The blue circles display the observed data and the lines represent the models corresponding to values of $\alpha$ as indicated in the legend. The $V$-band photometric data are shown with grey triangles in the background. The vertical dashed lines display the boundaries between the cycles. The vertical dotted lines show the boundaries between outburst and quiescence phases. C4, O4 and Q4 stand for the fourth cycle, formation and quiescence phases, respectively. Bottom panel displays the goodness of the fit ($\chi^{2}_{\rm red}$) for each $\alpha$ value. For calculating the $\chi^{2}_{\rm red}$ the data were binned (green circles) to combine the observed points in the vicinity of epochs for which the model was computed.}
    \label{fig:variable_alpha_pol}
\end{figure}

Figure~\ref{fig:variable_alpha_pol} displays the results of this test for the polarimetric data and confirms that larger values of $\alpha$ of about 0.17 (with reduced $\chi^{2}$, $\chi^{2}_{\rm red}=11$) enhance the fit significantly. Recall that the optimum value found for the $\alpha$ parameter for Q4 by fitting the $V$-band data was 0.11 \citepalias[][with $\chi^{2}_{\rm red}=69$; also see Figure~\ref{fig:summary}]{ghoreyshi2018}. With a greater $\alpha$, the  rate of polarimetric variation of the model is larger due to the increased dissipation rate, matching the lower data points that were not reproduced by the original model (see Figure~\ref{fig:polarimetry_model}).

\begin{figure}
    \begin{minipage}{\linewidth}
        \centering
        {\includegraphics[width=\columnwidth]{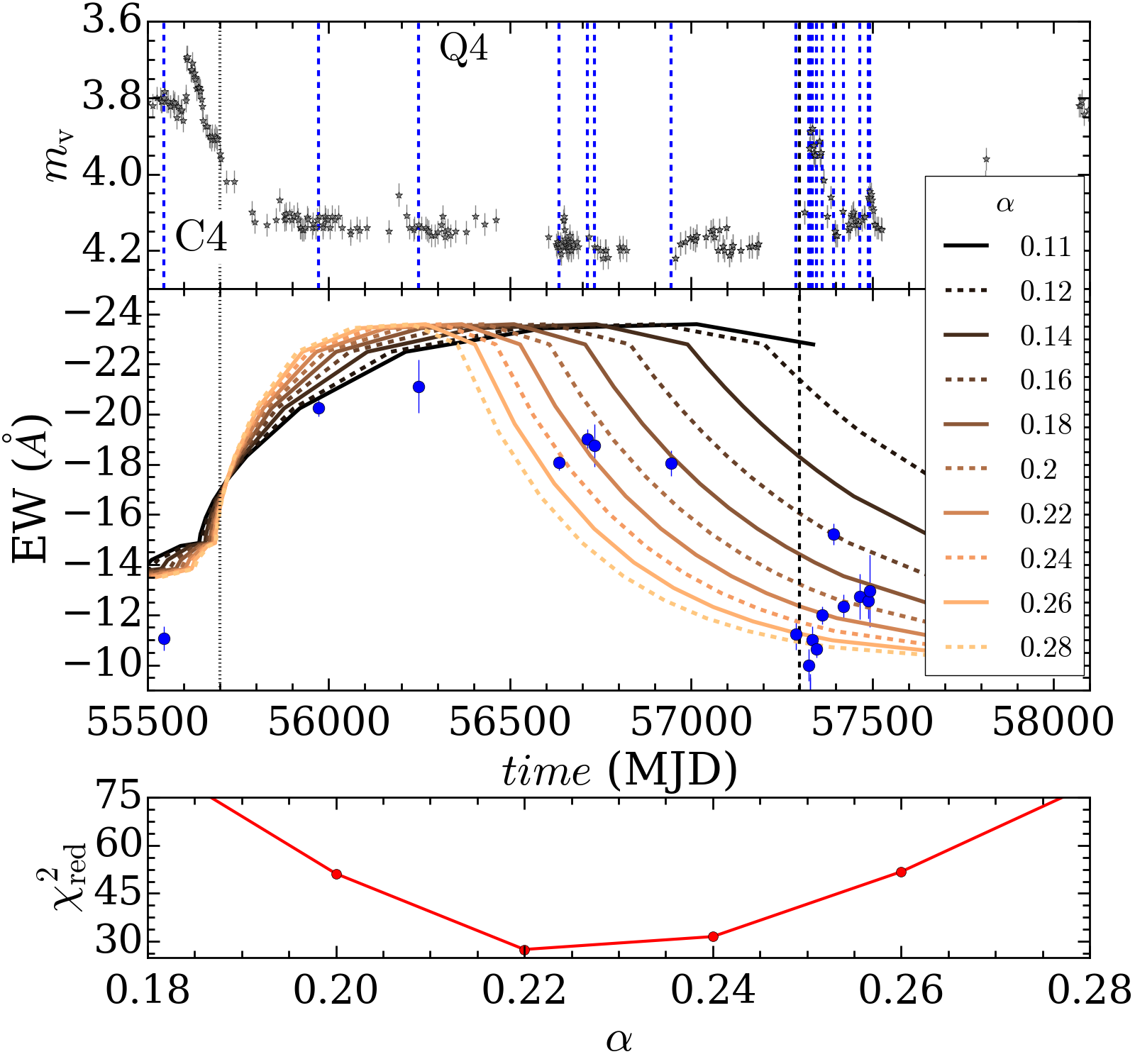}}
    \end{minipage}
    \caption{
   Comparison between the models with a range of $\alpha$ values and the observed H$\alpha$ EW. The upper panel shows the observed $V$-band photometric data, and the middle panel displays the EW for each model with the individual $\alpha$ parameter. The blue circles display the observed data and the lines represent the models with different $\alpha$ parameter as indicated in the legend. C4 and Q4 stand for the fourth cycle and quiescence phase, respectively. The vertical dashed lines display the boundaries between the cycles. The vertical dotted lines show the boundaries between outburst and quiescence phases. The lower panel demonstrates the goodness of the fit ($\chi^{2}_{\rm red}$) for each $\alpha$ value.}
    \label{fig:variable_alpha_sp4}
\end{figure}

Figure~\ref{fig:variable_alpha_sp4} confirms that an even larger value of $\alpha$ such as 0.22, leading to quicker disk evolution, is required for a better agreement between the model and the data. With a larger value for $\alpha$, the EW rises fast enough to match the data and later, at the middle of the dissipation phase, starts to drop simultaneously with the H$\alpha$ EW data. 

\begin{figure}
    \begin{minipage}{\linewidth}
        \centering
        {\includegraphics[width=\columnwidth]{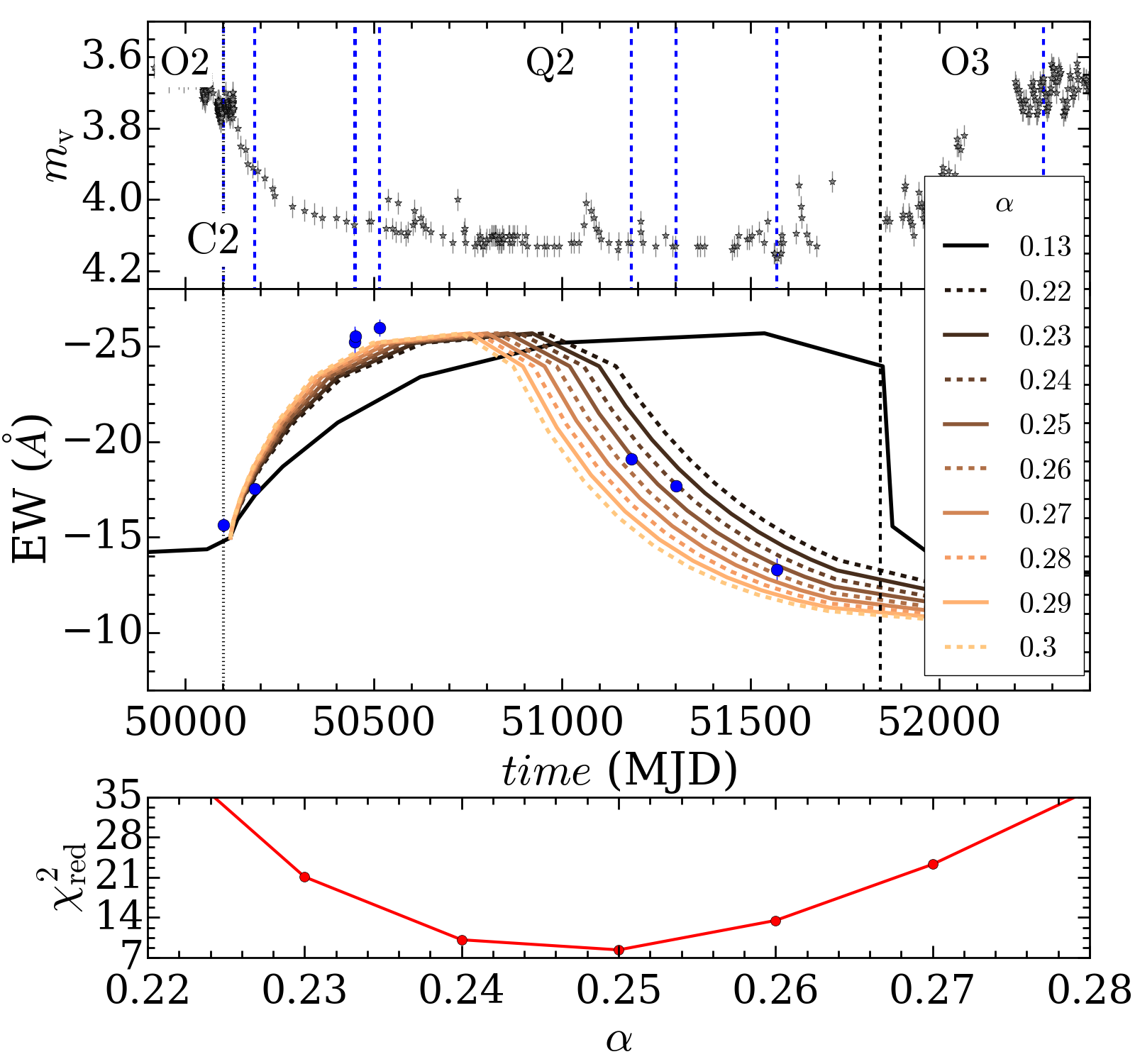}}
    \end{minipage}
    \caption{Same as Figure~\ref{fig:variable_alpha_sp4} for C2.}
    \label{fig:variable_alpha_sp2}
\end{figure}

Since the EW data for the H$\alpha$ line in C4 are sparse, we repeat this test for Q2. Figure~\ref{fig:variable_alpha_sp2} demonstrates a result similar to Q2: a larger value of $\alpha$ of about 0.25 with $\chi^{2}_{\rm red}\approx7.0$ (rather than the original value of 0.13 with $\chi^{2}_{\rm red}\approx130$ determined from $V$-band photometric modeling) provides much more consistency between the data and the model. These tests indicate a common pattern: larger values of $\alpha$ than what was obtained for the $V$-band lightcurve are required to match the observed rate of variations for the polarization (Q4) and H$\alpha$ EW (both Q2 and Q4). Therefore, recalling that the polarization and H$\alpha$ probe a radial extent of the disk of about 1.5 -- 5 times larger than the $V$-band continuum, respectively, this may suggest that the $\alpha$ parameter grows with distance from the star.

\begin{figure}
    \begin{minipage}{\linewidth}
        \centering
        {\includegraphics[width=\columnwidth]{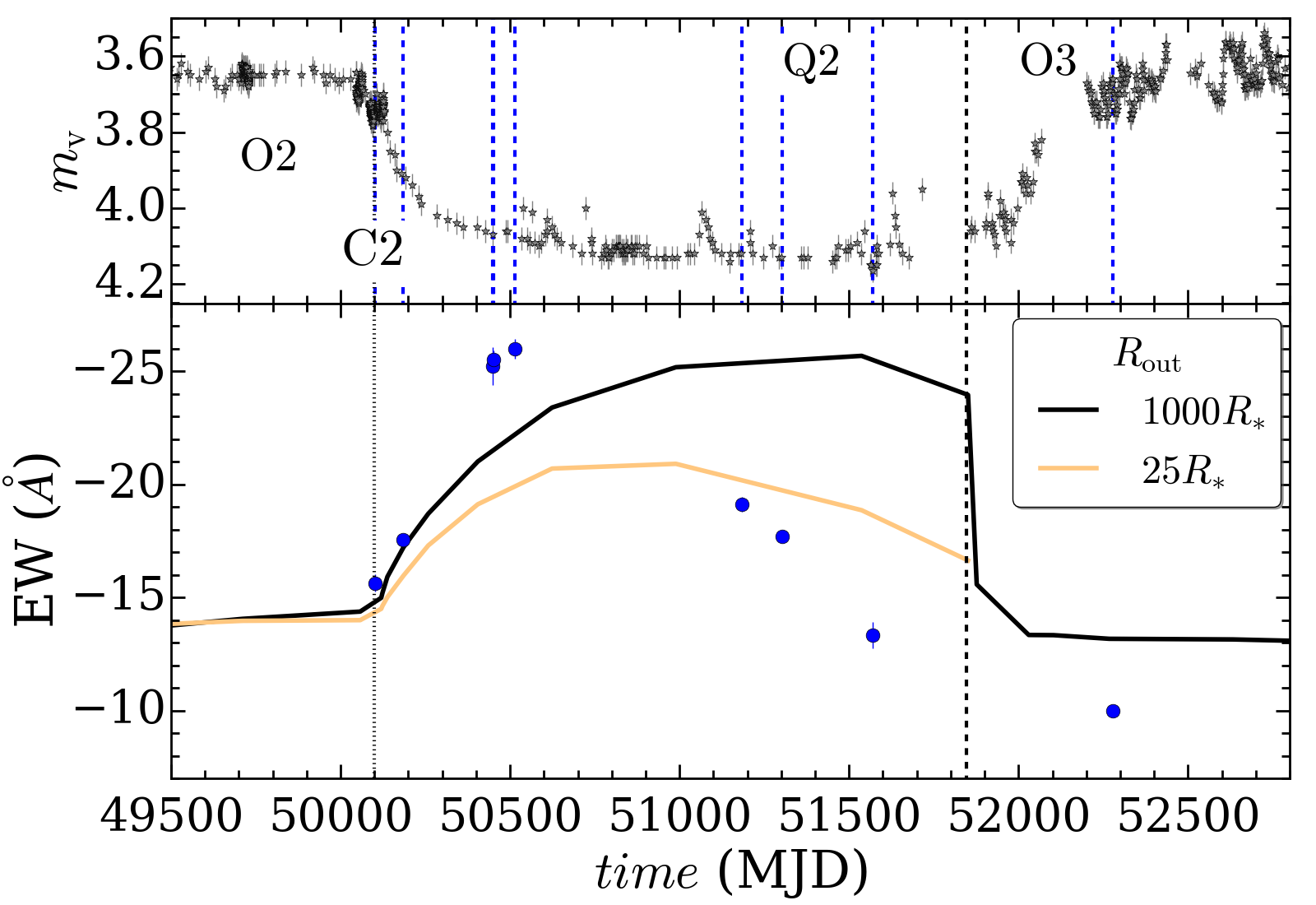}}
    \end{minipage}
    \caption{Comparison between the model presented in \citetalias{ghoreyshi2018} and the truncated disk model. The top panel illustrates the observed $V$-band photometric data, and the bottom panel displays the EW for each model. The blue circles demonstrate the observed data and the lines represent the models with different disk size as shown in the legend. C$i$, O$i$ and Q$i$ stand for cycle, outburst and quiescence phases, respectively, where $i$ is the cycle number. The vertical dashed lines display the boundaries between the cycles. The vertical dotted lines show the boundaries between outburst and quiescence phases.}
    \label{fig:truncation}
\end{figure}

For the second test (disk truncation), we calculated the H$\alpha$ EW during Q2 for a disk with $R_{\rm out}=25\,R_{\rm eq}$. The result for the H$\alpha$ line is shown in Figure~\ref{fig:truncation} with $\chi^{2}_{\rm red}\approx210$ and is compared with the result of \citetalias{ghoreyshi2018} that uses $R_{\rm out}= 1000\,R_{\rm eq}$ (the same model as $\alpha=0.13$ in Figure~\ref{fig:variable_alpha_sp2} with $\chi^{2}_{\rm red}\approx130$). The mass reservoir \citep[see][for details]{rimulo2018} for the smaller disk is reduced, therefore, the smaller disk dissipates faster and, consequently, the EW drops faster. On the other hand, the smaller disk mass produces a smaller H$\alpha$ strength, as shown in Figure~\ref{fig:truncation}. Thus, although the truncated disk hypothesis seems a viable solution for increasing the rate of dissipation, it creates another problem, namely an H$\alpha$ emission that is too small. It is worth mentioning that \cite{harmanec1998} and \cite{stefl2003a} did not find any evidence of binarity.  However, if the orbital plane and the circumstellar disk plane are about the same, the radial velocity signature of a companion is very difficult to find owing to the small inclination angle, especially if it is a subluminous star. The results discussed above are unchanged regardless of the cycle, because the trends seen in the EW curve of all cycles are similar.

Finally, it is worth discussing an effect that was not included in our models but could affect the results by increasing the disk dissipation rate. \cite{kee2016a} showed that radiation forces, especially for the very hot O-type stars with strong winds, can ablate the entire disk in timescales of the order of days to years. They suggest that this is the reason why disks are not commonly observed surrounding these types of stars. They also showed that for a B2-type star it would take a couple of months to destroy an optically thin, low density disk. However, \cite{kee2018b} concluded that, for a more massive optically thick disk (like $\omega$ CMa) this effect would decrease the ablation rate by only 30\% or less. Therefore, although the ablation does not seem adequate for the disk dissipation timescale of $\omega$ CMa ($\approx$ 4.5 to 6.5 years), it is possible that radiative ablation plays a role in Be disk dynamics, but the extent of this role remains to be determined and the most significant ablation would occur for the innermost disk. In principle, if radiative ablation is important in the case of $\omega$ CMa, the values of $\alpha$ for the dissipation phases quoted in \citetalias{ghoreyshi2018} and Section~\ref{sect:alternative} of this paper would represent upper limits. However, since this mechanism does not seem to be strong enough in the outer parts of the disk, it cannot help the problem of slow dissipation of the disk seen in our models. Future work, combining viscosity and ablation, is necessary to properly address this problem.


\section{Conclusions}
\label{sect:conclusions}

We use the VDD model to study the observed data from a range of wavelengths and techniques for the Be star, $\omega$ CMa, in a dynamical fashion. We adopt the same model presented in \citetalias{ghoreyshi2018} that was used to study the $V$-band continuum emission of this star. In this work, we compare model predictions with a range of different observations (multi-band photometry, spectroscopy, and polarimetry). Since different wavelengths originate from various parts of the disk (Figure~1 of \citealt{carciofi2011}), this method is a solid test for the VDD model.

The results were mostly positive: qualitative and even quantitative agreements were found, but in some cases important differences could be noted. We see the best agreement for the visible photometric data, but for the IR band the differences are more significant. This makes sense because the model parameters used here are the same as \citetalias{ghoreyshi2018} that were used for modeling the $V$-band originating mostly from the inner regions of the disk ($\lesssim 2\,R_{\rm eq}$) while the IR photometric data (e.g., {\it JHKL} and $Q$1, $Q$3) come from a much larger volume of the disk ($\lesssim 10\,R_{\rm eq}$). The models predicted larger IR excesses than observed, indicating that the disk mass is likely overestimated.

The average polarization level was fitted acceptably by our models, but the model rate of dissipation during quiescence was too slow. One important point to stress is that the observations showed a much slower decline rate of the polarization than the $V$-band light curve. This slower decay indicates that the polarization originates in a larger radial extent than the $V$ band continuum, 
which is consistent with our model predictions.

Although our model could fit the spectroscopic data qualitatively, we find, similarly to the polarization level, that the rate of EW decay during quiescence is too slow. This seems to point to the fact that in the original \citetalias{ghoreyshi2018} model the rate of density variation in the outer disk is too small, meaning that, during quiescence and at larger distances from the star, the disk is not being drained fast enough. 
Two tests were conducted to look for possible remedies to this issue.

In the first test, we experimented with models with increased values of $\alpha$. These models produced an EW curve very similar to the observed one, specially for Q4. Given that lower values of $\alpha$ are required to match the $V$-band light curve, as in \citetalias{ghoreyshi2018}, these results hint at the possibility of a radially increasing $\alpha$ in Be star disks. 

As a second alternate scenario we considered the effect of truncating the disk by an unresolved binary companion that could potentially decrease the density in the outer part of the disk. This effect was considered by changing the outer radius of the simulation from $1000\,R_{\rm eq}$ to $25\,R_{\rm eq}$. While the rate of line strength variation approached the observed data, truncating the disk created the undesired effect of reducing the line emission.

It is important to note that the above results could be interpreted in a different way. The discrepancies above might simply be the result that the $V$-band photometry alone cannot fully constrain the disk at all radii. As the \citetalias{ghoreyshi2018} model likely suffers from degeneracies, it is possible that different model parameters (e.g., different $\dot J$ and values of the $\alpha$ parameter) might be able to explain the full set of observations, without the need to resort to the variable $\alpha$ or truncated disk scenarios. This possibility will be explored in future models. 

Also, it is worth mentioning that line-driven ablation may play a role helping the disk to dissipate faster \citep{kee2016a, kee2018b, kee2018a}. This means that if this effect is important, the values we find for the $\alpha$ parameter are upper limits. However, ablation affects mostly the inner disk. Therefore, its role, if any, should be more noticeable in the observables that are more sensitive to the inner disk variations, e.g., $V$-band photometry and polarimetry. In this regard, including ablation in the model might exacerbate the mismatch between observations and model concerning the rate of polarimetric and spectroscopic variations during quiescence.

Finally, it is worth discussing the instability of the disk of $\omega$ CMa. As an early-type Be star, it is more likely that $\omega$ CMa possesses an unstable disk in comparison to the late-type Be stars \citep{labadie2018} and all our multi-technique data confirm this statement, showing variations on timescales of few days to several years. The origins of this instability may be caused by one or more of the following mechanisms: discrete mass-loss events caused by the nonlinear coupling of multiple NRP modes \citep[e.g.,][]{baade2016}, fast rotation (e.g., \citealt{rivinius2013b} reports variations of the width of photospheric lines likely linked to changes in the rotation rate of the surface layers), and ablation \citep[e.g.,][]{kee2016a}.

In Section~\ref{sect:alternative} we discussed the possible existence of an undetected binary companion and its effect on disk size. Investigating the long-baseline interferometric data of $\omega$ CMa may help us to have a better understanding of the morphology of the disk and of the possible existence of a companion object. In the future, we also plan to extend our analysis to include interferometric data and long wavelength (radio) photometry and Balmer decrement variations in order to continue to explore the limits of the VDD model.


\section*{acknowledgements}

We would like to thank the anonymous referee for her/his constructive criticism, careful reading, thoughtful and insightful comments and time and efforts towards improving our manuscript.

This work made use of the computing facilities of the Laboratory of Astroinformatics (IAG/USP, NAT/Unicsul), whose purchase was made possible by the Brazilian agency FAPESP (grant 2009/54006-4) and the INCT-A. M.R.G. acknowledges the support from CAPES PROEX Programa Astronomia and the grant awarded by the Western University Postdoctoral Fellowship Program (WPFP). A.C.C acknowledges support from CNPq (grant 311446/2019-1) and FAPESP (grant 2018/04055-8). C.E.J. acknowledges the Natural Sciences and Engineering Research Council of Canada for the financial support. D.M.F acknowledges Supports by the international Gemini Observatory, a program of NSF’s NOIRLab, which is managed by the Association of Universities for Research in Astronomy (AURA) under a cooperative agreement with the National Science Foundation, on behalf of the Gemini partnership of Argentina, Brazil, Canada, Chile, the Republic of Korea, and the United States of America. 

This research used the observations collected at the European Organization for Astronomical Research in the Southern Hemisphere under ESO programs 68.D-0095(A), 68.D-0280(A), 69.D-0381(A), 74.D-0240(A), 75.D-0507(A), 82.A-9202(A), 82.A-9208(A), 82.A-9209(A), 268.D-5751(A), and 282.D-5014(B).

Some of the data presented in this paper were obtained from the Mikulski Archive for Space Telescopes (MAST). STScI is operated by the Association of Universities for Research in Astronomy, Inc., under NASA contract NAS5-26555. Support for MAST for non-HST data is provided by the NASA Office of Space Science via grant NNX13AC07G and by other grants and contracts.

This work has made use of data from the European Space Agency (ESA) mission
{\it Gaia} (\url{https://www.cosmos.esa.int/gaia}), processed by the {\it Gaia}
Data Processing and Analysis Consortium (DPAC,
\url{https://www.cosmos.esa.int/web/gaia/dpac/consortium}). Funding for the DPAC
has been provided by national institutions, in particular the institutions
participating in the {\it Gaia} Multilateral Agreement.

The IR-band data were provided by Juan Fabregat Llueca and they were obtained at the South African Astronomical Observatory (SAAO), and at the Teide Observatory (Tenerife, Spain).

We thank Wagner J.B. Corradi from Universidade Federal de Minas Gerais for providing part of the polarimetric data and Daniel Bednarski from Universidade de S\~{a}o Paulo for providing the polarimetric data of HD56876.

Also, this research has made use of the SIMBAD database and VizieR catalogue access tool, operated at CDS, Strasbourg, France.

\bibliography{bibliography}

\clearpage
\pagebreak

\appendix

\section{Observational Log}
\label{ap:obs_log}

\begin{table}[!b]
\begin{center}
\caption{Spectroscopic Data Logs}
\begin{tabular}{@{}cccc}
\hline
\hline
Reference & Number of Points & Time Coverage (MJD) & Wavelength ({\AA})\\
\hline
CES & 2 & 52659 -- 52660 & 4184 -- 7314\\
ESPaDOnS & 16 & 55971 -- 55971 & 3696 -- 8868\\
{\sc feros} & 444 & 52277 -- 54822 & 3527 -- 9215\\
{\sc heros} & 435 & 50102 -- 51301 & 3438 -- 8629\\
IUE & 12 & 43833 -- 44975 & 1000 -- 3200\\
Lhires & 25 & 54083 -- 57465 & 6511 -- 6610\\
Ondrejov & 7 & 53060 -- 56737 & 6258 -- 6770\\
OPD & 8 & 56636 -- 57645 & 4118 -- 9183\\
PHOENIX & 10 & 54776 -- 55311 & 21604 -- 21700\\
Ritter & 20 & 57329 -- 57496 & 6471 -- 6634\\
UVES & 141 & 54784 -- 54913 & 3055 -- 10426\\  
\hline
\end{tabular}
\label{table:spec_data_log}
\end{center}
\end{table}


\begin{table}[!b]
\begin{center}
\caption{Polarimetric Data Logs, observed by OPD}
\begin{tabular}{@{}ccc}
\hline
\hline
Reference & Number of Points & Time Coverage (MJD) \\
\hline
$U$ Filter & 3 & 55497 -- 56050\\ 
$B$ Filter & 39 & 54765 -- 57624\\ 
$V$ Filter & 81 & 54505 -- 57626\\ 
$R$ Filter & 37 & 54975 -- 57624\\ 
$I$ Filter & 39 & 54975 -- 57624\\ 
\hline
\end{tabular}
\label{table:pol_data_log}
\end{center}
\end{table}

\pagebreak

\section{Additional Polarimetric data}
\label{ap:polari}


\begin{figure}[!b]

\begin{minipage}{.5\linewidth}
\centering
\subfloat[]{\label{main:a}\includegraphics[scale=.2]{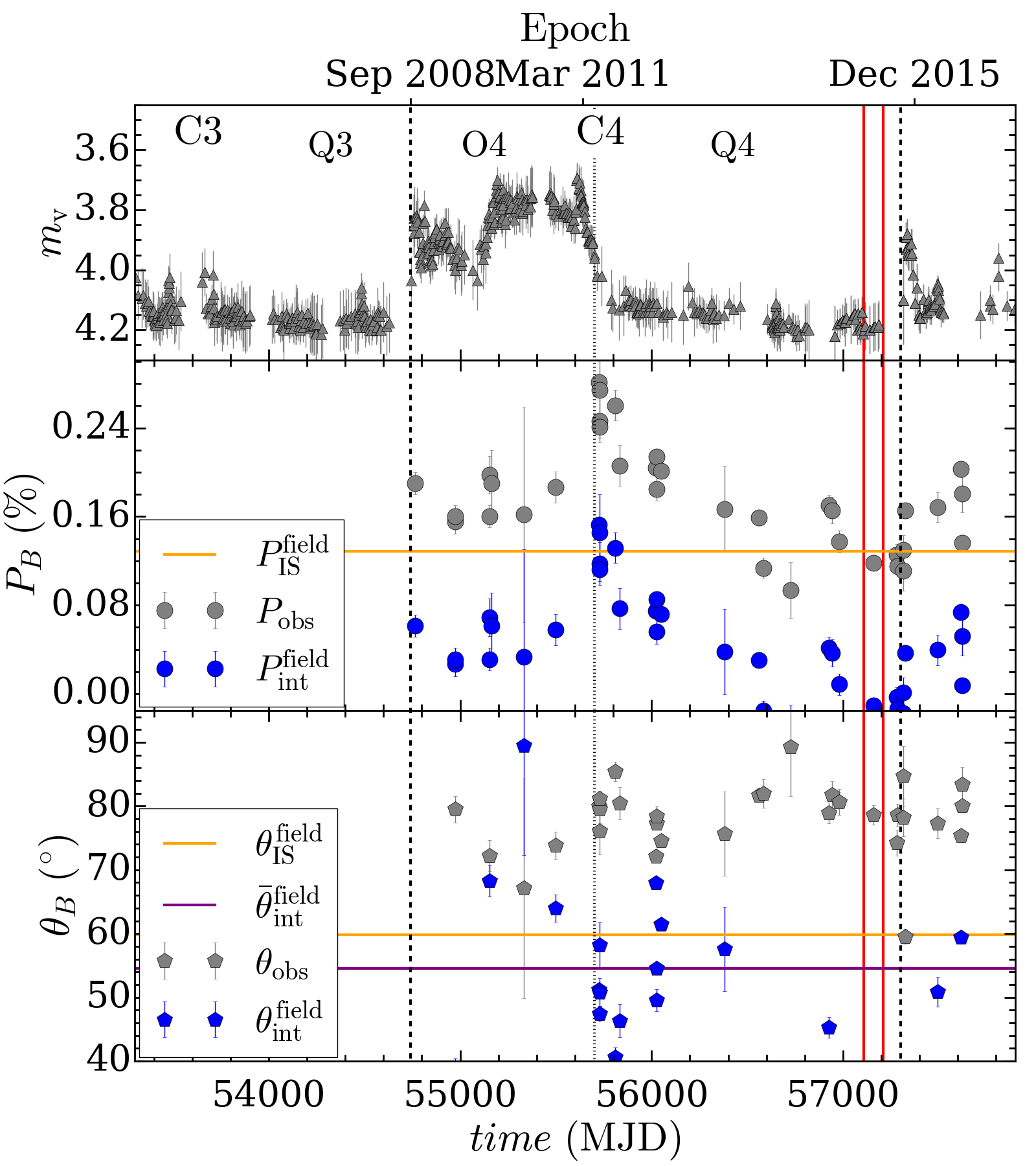}}
\end{minipage}%
\begin{minipage}{.5\linewidth}
\centering
\subfloat[]{\label{main:b}\includegraphics[scale=.2]{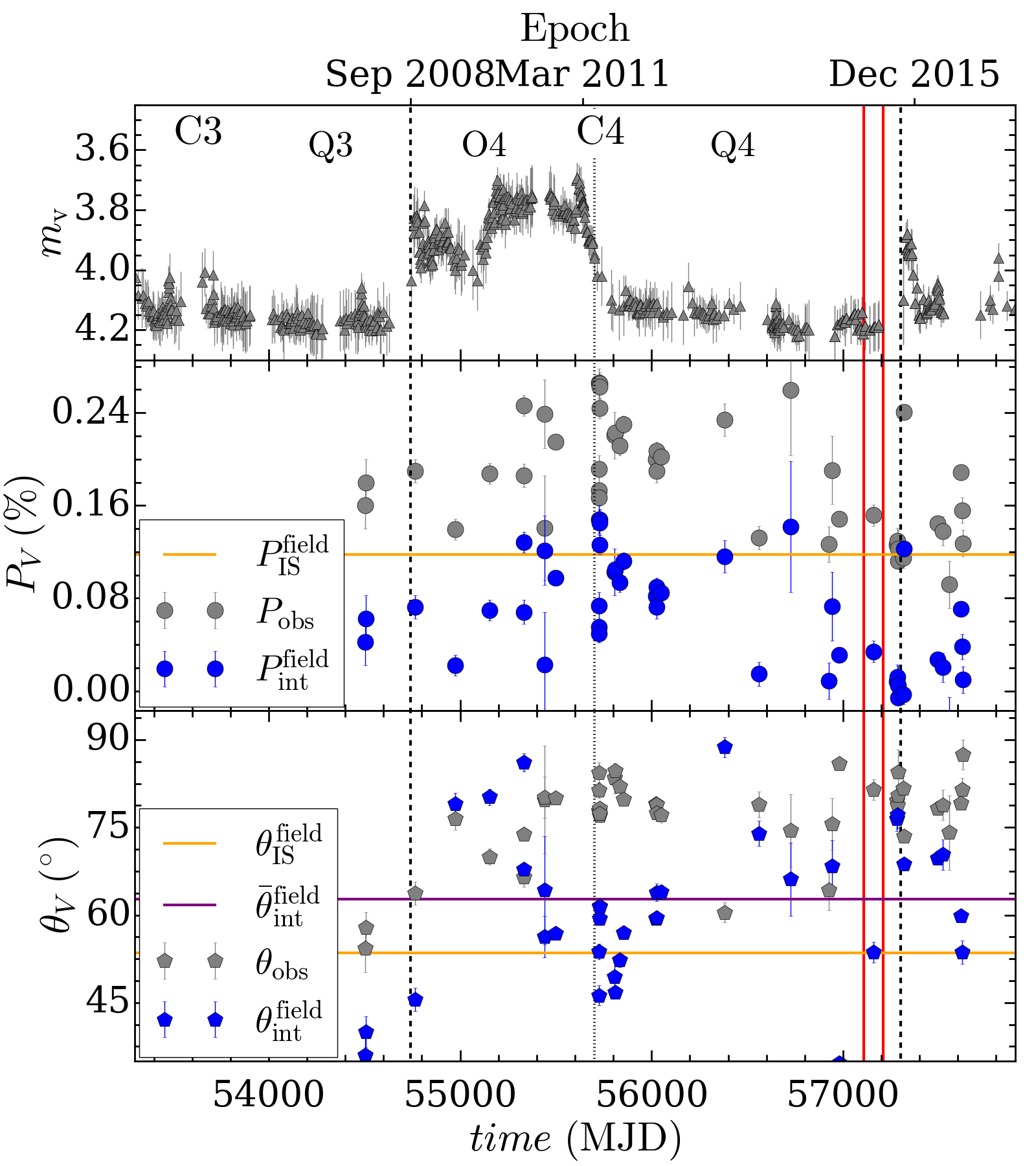}}
\end{minipage}
\par\medskip
\begin{minipage}{.5\linewidth}
\centering
\subfloat[]{\label{main:c}\includegraphics[scale=.2]{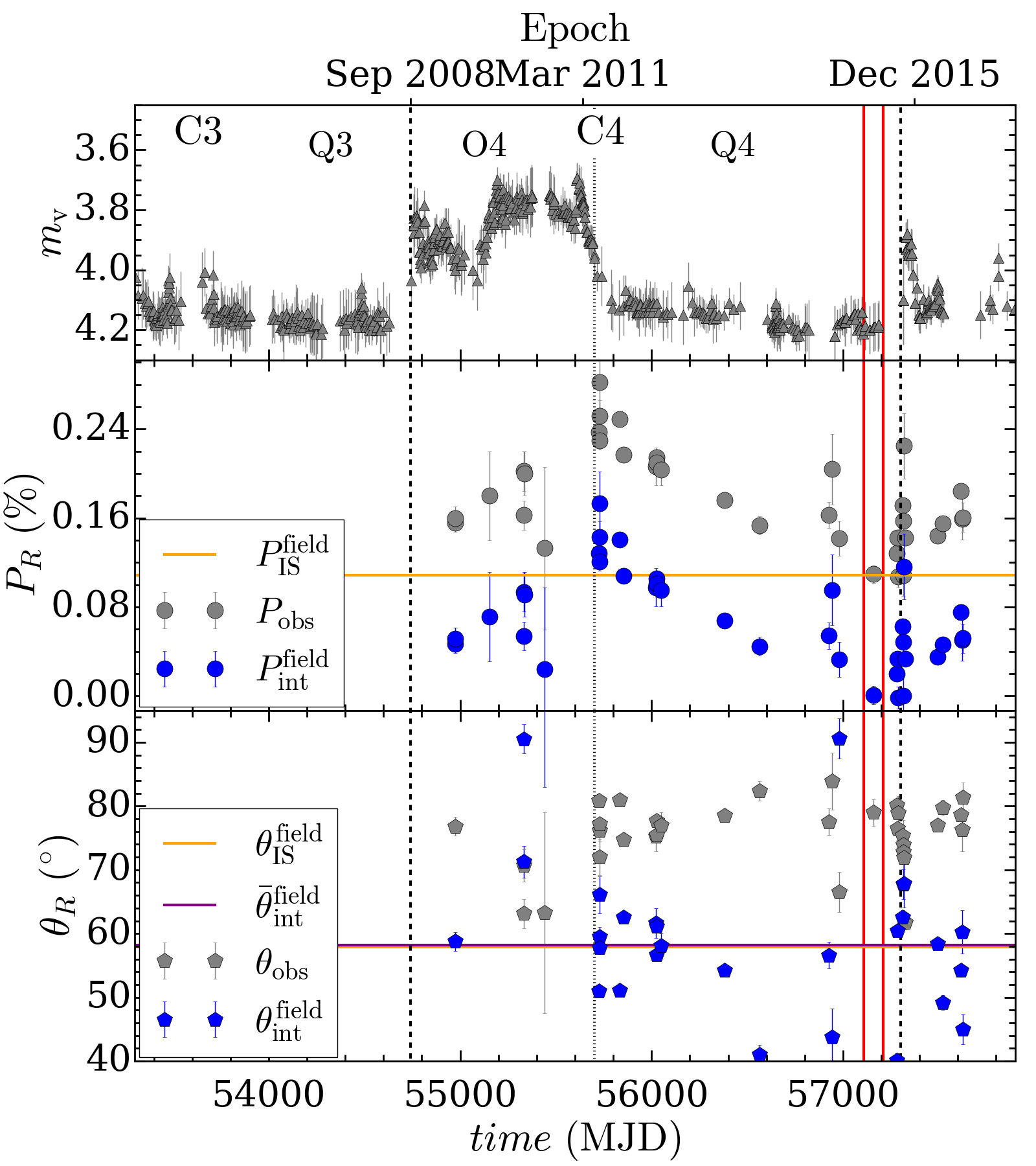}}
\end{minipage}%
\begin{minipage}{.5\linewidth}
\centering
\subfloat[]{\label{main:d}\includegraphics[scale=.2]{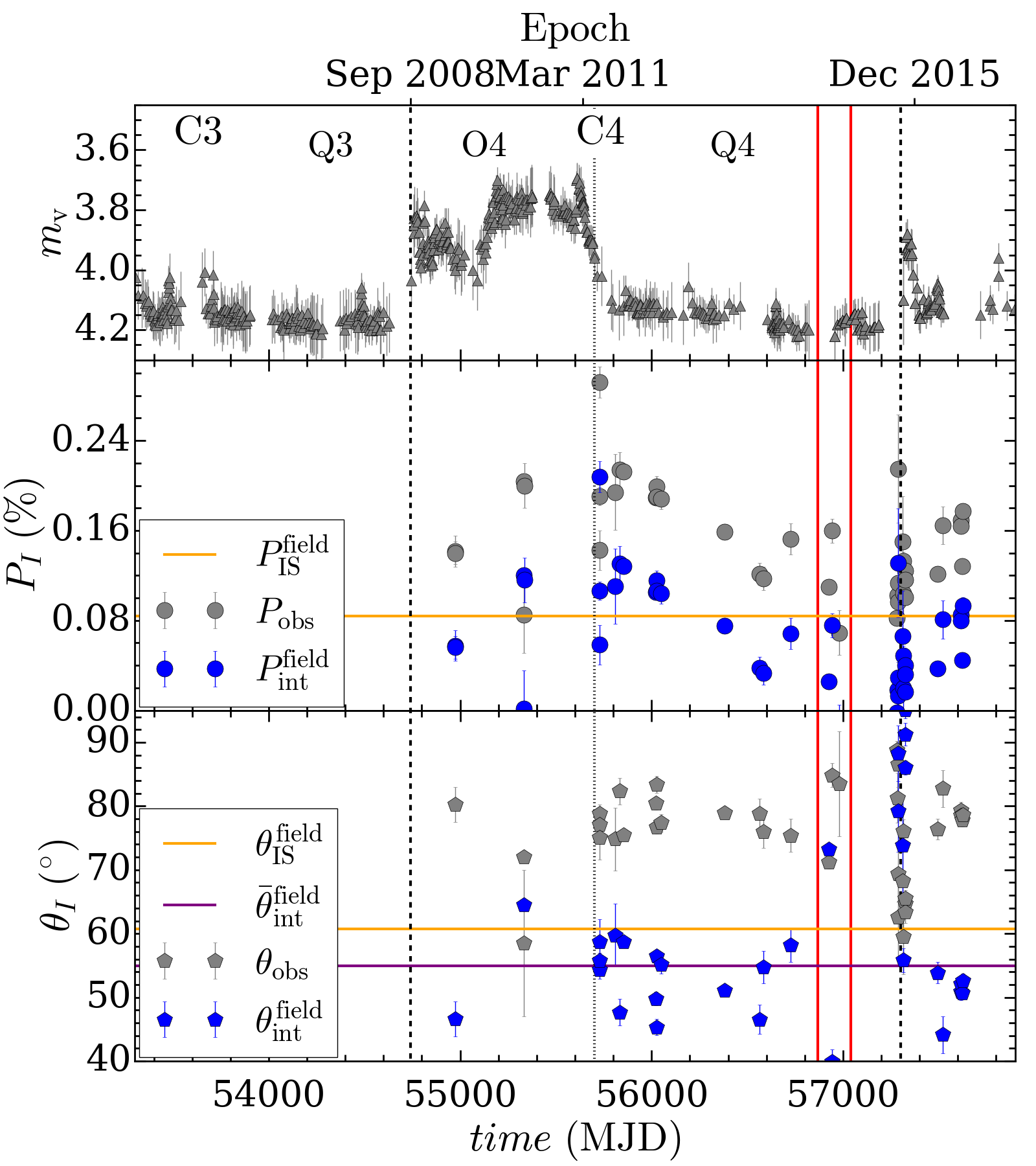}}
\end{minipage}
\caption{Same as Figure~\ref{fig:pol_v_corrected} for (a) $B$, (b) $V$, (c) $R$ and (d) $I$ filters. Also, the position angle for each filter is shown. The observed (grey pentagons) and intrinsic polarization angles (blue pentagons) of $\omega$ CMa. The $\theta_{\rm IS}^{\rm field}$ and the average value of the intrinsic polarization angle are shown with orange and purple horizontal lines, respectively. The vertical dashed lines display the boundaries between the cycles. The vertical dotted lines show the boundaries between outburst and quiescence phases.}
\label{fig:pol_bvri_corrected}
\end{figure}

 \pagebreak

\section{Model fits for H$\gamma$ and H$\delta$ lines}
\label{ap:spectro}

\begin{figure*}[!b]
    \begin{minipage}{0.5\linewidth}
        \centering
        \subfloat[H$\gamma$]{\includegraphics[width=1.0\linewidth]{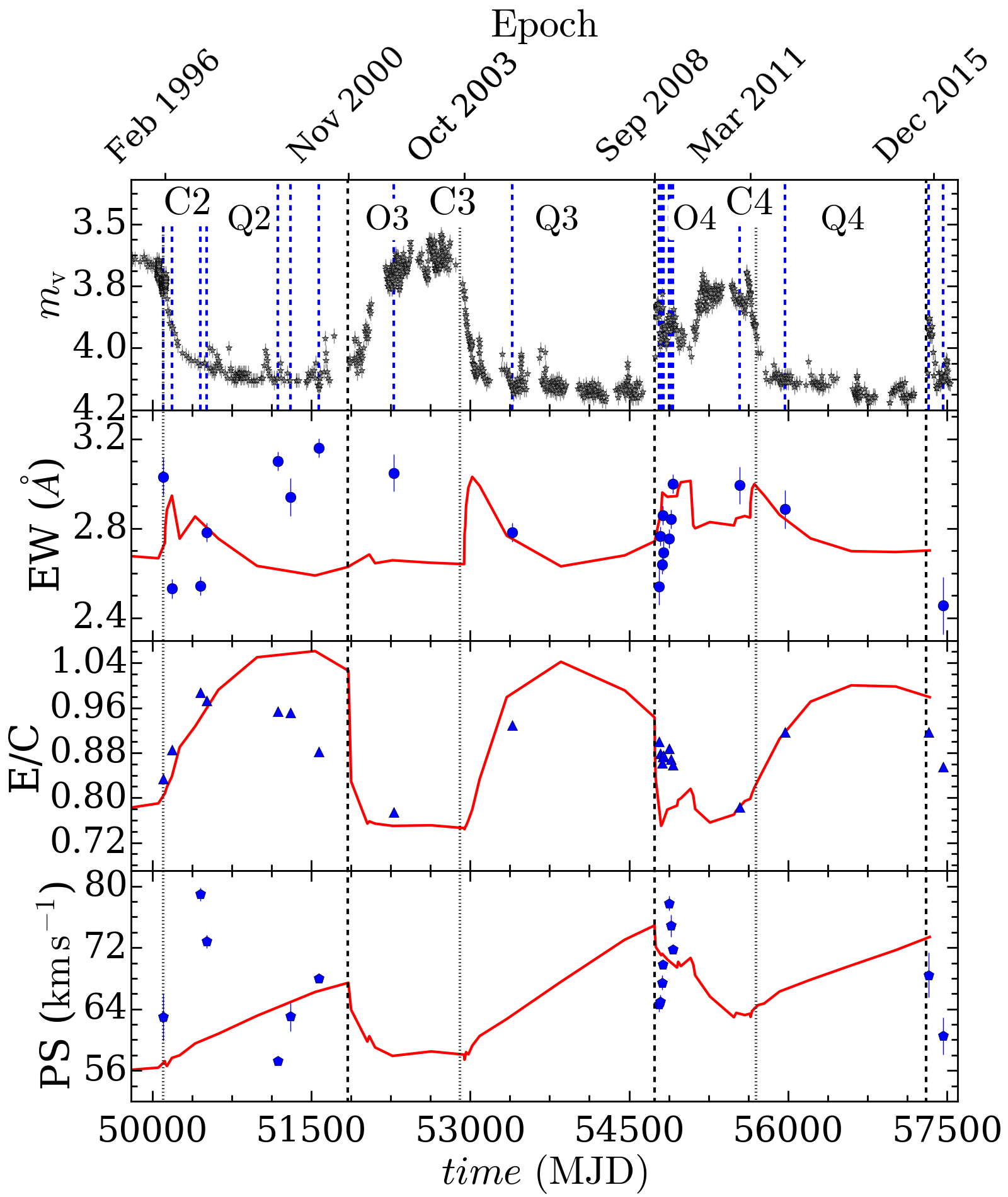}}
    \end{minipage}%
    \begin{minipage}{0.5\linewidth}
        \centering
        \subfloat[H$\delta$]{\includegraphics[width=1.0\linewidth]{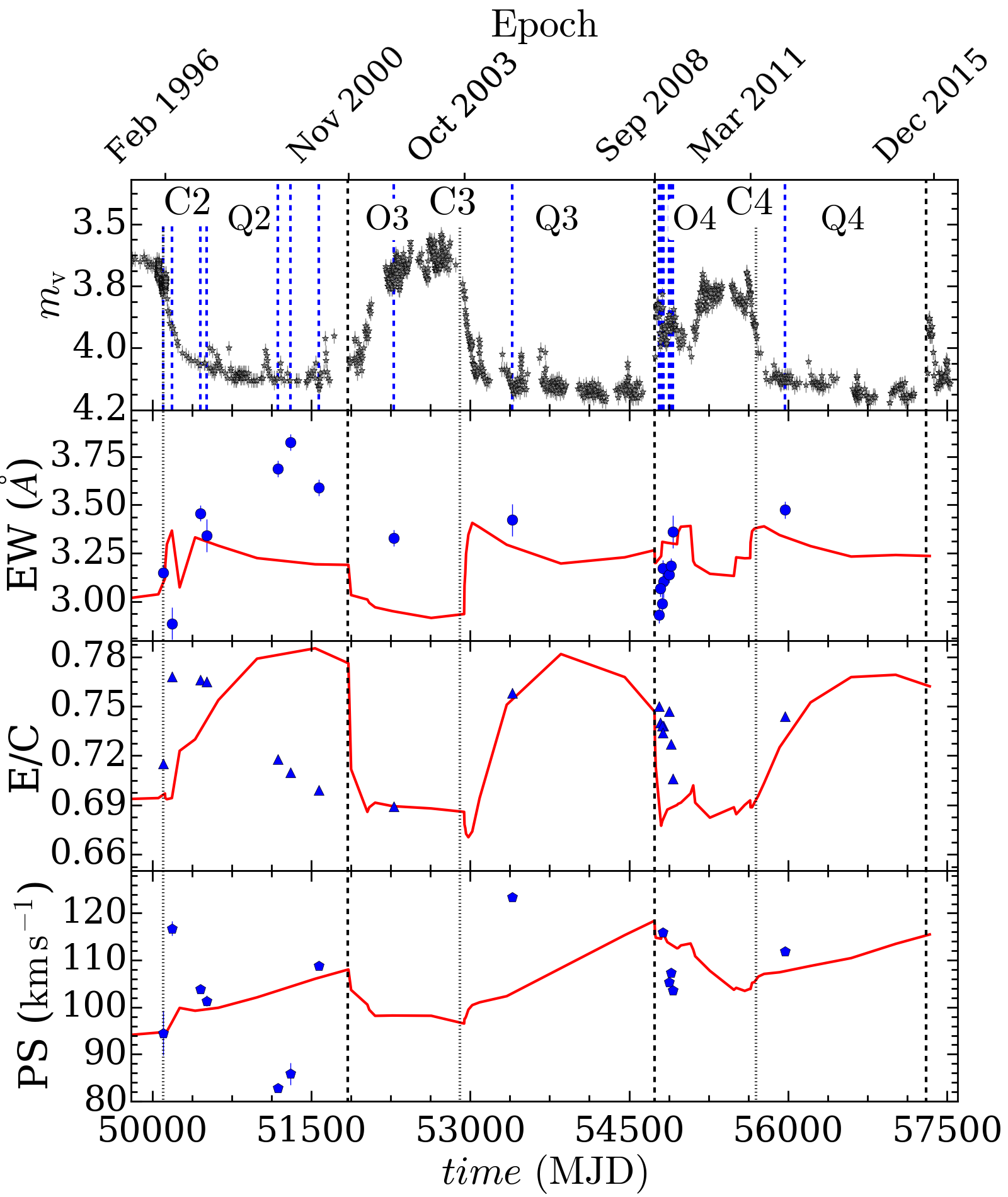}}
    \end{minipage}
    \caption{Same as Figure~\ref{fig:spectroscopy_model_balmer} for (a) H$\gamma$ and (b) H$\delta$ lines.}
    \label{ap:spectroscopy_model_balmer_2}
\end{figure*}


\end{document}